\documentclass[twocolumn,10pt]{IEEEtran}
\usepackage{ifpdf, flushend,subfigure}
\ifCLASSINFOpdf
  \usepackage[pdftex]{graphicx}
  \graphicspath{{../pdf/}{../jpeg/}}
  \DeclareGraphicsExtensions{.pdf,.jpeg,.png}
\else
  \usepackage[dvips]{graphicx}
  \graphicspath{{../eps/}}
  \DeclareGraphicsExtensions{.eps}
\fi
\usepackage{graphicx}
\usepackage{epstopdf}
\graphicspath{{../}}
\usepackage{multirow}
\usepackage{float}
\usepackage[cmex10]{amsmath}
\usepackage {amssymb}
\usepackage{array}
\usepackage{mdwmath}
\usepackage{mdwtab}
\usepackage{eqparbox}
\usepackage{nomencl}
\usepackage{cite}
\usepackage{algorithm}
\usepackage{algorithmic}
\usepackage{makeidx}
\usepackage{ifthen}
\usepackage{subfigure}
\usepackage{caption}
\usepackage{pdflscape}
\usepackage{hyperref}
\newcommand{\beq}{\begin{equation}}
\newcommand{\eeq}{\end{equation}}
\newcommand{\beqn}{\begin{eqnarray}}
\newcommand{\eeqn}{\end{eqnarray}}
\newcommand{\beqno}{\begin{eqnarray*}}
\newcommand{\eeqno}{\end{eqnarray*}}
\newcommand{\bma}{\begin{displaymath}}
\newcommand{\ema}{\end{displaymath}}
\newcommand{\bnu}{\begin{enumerate}}
\newcommand{\enu}{\end{enumerate}}
\newcommand{\bce}{\begin{center}}
\newcommand{\ece}{\end{center}}
\newcommand{\btb}{\begin{tabular}}
\newcommand{\etb}{\end{tabular}}

\newcommand{\floor}[1]{\left\lfloor #1 \right\rfloor}

\begin{document}

\title{Two-Stage Robust Edge Service Placement and Sizing under Demand Uncertainty}

\author{\IEEEauthorblockN{Duong Tung Nguyen,  Hieu Trung Nguyen, Ni Trieu,  and Vijay  K. Bhargava} } 





\maketitle

\begin{abstract}
Edge computing has emerged as a key technology to reduce network traffic, improve user experience, and enable various Internet of Things   applications.
From the perspective of a service provider (SP), how to jointly optimize the service placement, sizing, and workload allocation decisions is an important and challenging problem, which becomes even more complicated when considering  demand uncertainty.
To this end, we propose a novel two-stage adaptive robust optimization framework to help the SP optimally determine the locations for installing their service (i.e., placement) and  the amount of computing resource to purchase from each location (i.e., sizing).
The service placement and sizing solution of the proposed model can hedge against any possible realization within the   uncertainty set of traffic demand.
 Given the  first-stage robust solution, the optimal resource and workload allocation decisions are computed in the second-stage after the uncertainty is revealed.
To solve the  two-stage model, 
in this paper, we present an iterative solution by employing  the column-and-constraint generation method 
that decomposes the underlying problem into a master problem and a
max-min subproblem associated with the second stage. 
Extensive numerical results are shown to illustrate the effectiveness of the proposed  two-stage  robust optimization model. 
\end{abstract}

\begin{IEEEkeywords}
Edge computing, service placement, workload allocation, adaptive robust optimization, demand uncertainty.
\end{IEEEkeywords}

\printnomenclature


\section{Introduction}
Edge computing (EC) has been proposed  to augment the traditional cloud computing model to  meet the soaring traffic demand and accommodate diverse requirements of various services and systems in future networks, such as embedded artificial intelligence (AI), 5G wireless systems,  virtual/augmented reality (VR/AR), 
and tactile Internet  \cite{mchi16,wshi16}.
By distributing storage, computing, control, and networking resources closer to the network edge, EC
offers  remarkable advantages and capabilities, including local data processing and analytics, localized services,  edge caching,  edge resource pooling and sharing,   and improved privacy and  security  \cite{duong1,duong2}.
Also, EC is   a key enabler for
ultra-reliable low-latency applications.

To enhance  user experience and reduce bandwidth usage,
content/application/service providers (e.g., AR/VR companies, Google, Netflix, Facebook,  Uber, Apple,   and other OTT providers) can  proactively install their  applications, especially latency-sensitive and/or data-intensive ones  such as AR/VR,  cloud gaming, and video analytics,  onto selected edge nodes (EN)  in proximity of their users.
Therefore, in addition to local execution on end-devices  and remote processing in public clouds or their private data centers (DC), 
the SPs can offload their  tasks to  edge servers. 
Besides SPs,  virtual network operators, vertical industries, enterprises,   and other third parties (e.g., schools, hospitals, malls, sensor networks) can also outsource their data and computation to  EC systems. 

\begin{figure}[h!]
	\centering
		\includegraphics[width=0.34\textwidth,height=0.14\textheight]{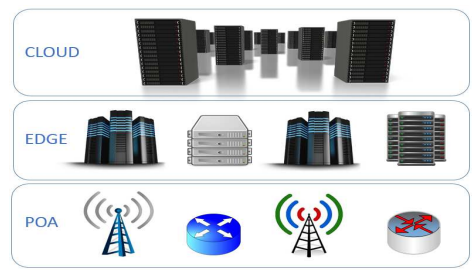}
			\caption{Edge Network Architecture}
	\label{fig:edge}
\end{figure}

Fig. \ref{fig:edge} depicts the new network architecture with an EC layer lying between the cloud and the aggregation layer.  In particular, the aggregation layer consists of numerous Points of Aggregation (POA), such as base stations (BS) and network routers/switches, which aggregate data and requests from  users, things, and sensors.
In practice, various sources (e.g., Telco edge clouds, telecom central offices,  servers at BSs,  PCs in research labs, micro DCs in campus buildings, malls,  and enterprises) can act as ENs \cite{mchi16,wshi16, duong1,duong2}.  Indeed,  an EN  can be co-located with a POA. For example, edge servers can be placed in cell sites or deployed near  routers in enterprise DCs.

Typically, a service request first arrives at a POA, then it will be routed to an EN or the remote cloud for processing. For instance,  with EC, when a user submits a Google Maps request or an Uber ride request,  the request can now be handled by an EN instead of going all the way to the remote servers of Google or Uber. Clearly, EC not only can help the SP drastically improve the service quality but also significantly lower network bandwidth consumption.

Despite  tremendous potential, EC is still in its infancy stage and many interesting open problems remain to be solved. 
In this work, we focus on the optimal edge service placement and workload allocation from the perspective of a SP (e.g., AR/VR, Pokémon Go, real-time translation, Uber,  Apple Siri, Amazon Alexa, Google Assistant,  Google Maps). Specifically, the SP needs to serve a large number of users/subscribers  located in different areas. The goal of the provider is  to minimize the total operating cost while maximizing the service quality. 
Here, we measure the quality of service (QoS) in terms of network delay.

In order to
 reduce the  delay between the users  and  computing nodes, the SP can provision the service on various distributed ENs. 
Then, each user  request can be processed by its closest ENs,
which have the service installed. It is easy to see that 
  placing the service on more ENs can lower the  overall delay, but it  also increases the SP's cost.
Specifically, when the service is available on more ENs, the network delay decreases because   requests in each service area can be routed to the ENs closer to them. On the other hand, the service placement cost increases when the service is installed on more ENs.
Hence, there is an inherent trade-off between the  operating cost of the SP and the overall network delay of the requests.

Furthermore, 
unlike the traditional cloud with virtually infinite capacity, ENs often have limited computational power \cite{mchi16,wshi16,duong1}. Additionally, in contrast to a small number of cloud DCs, there are numerous heterogeneous distributed ENs coming with different sizes and configurations. The  resource prices of the ENs can also be different due to various factors such as different hardware specifications,  electricity prices, location, reliability, reputation, and ownership. Thus, some ENs close to the users may not be chosen because of their higher  prices. 
As a result, selecting suitable ENs for service placement 
is a challenging task due to the heterogeneity of the ENs.

Besides the placement decisions, the SP also needs to decide the amount of  resource to buy from each selected EN. 
Given the service placement and sizing decisions, the provider will then decide how to allocate the traffic demand  in different areas to different ENs to minimize the overall network  delay.
To this end, when the traffic demand 
is known, we  formulate the  joint service placement, resource sizing, and workload allocation problem as a mixed integer linear program (MILP), which can be solved efficiently by leveraging the state-of-the-art MILP solvers.
The formulated problem aims to minimize the weighted sum of the operating cost of the SP and the total network delay of the user requests, while taking into account practical system design criteria such as resource capacity limits, budget constraint, and  delay preference.

This problem becomes more sophisticated when considering the demand uncertainty.
For example, in practice, the SP normally solves the deterministic MILP 
formulation using the forecast demand to find the optimal resource provisioning solution (i.e., placement and sizing). However, 
since the actual demand is unknown to the SP at the time of making decision, over-provisioning or under-provisioning may occur frequently, which is undesirable. Specifically, if the procured resources are excessive to serve the actual traffic demand most of the time, it leads to over-provisioning and unnecessarily high provisioning cost. On the other hand, if the procured resources are not sufficient to serve the actual demand most of the time, it leads to under-provisioning and may severely affect the quality of service (e.g., high latency, dropping requests).

A popular technique to deal with uncertainty is stochastic optimization, which has been successfully applied to many engineering problems, including cloud resource provisioning \cite{scha12, smir20, hbad20}. However, the stochastic optimization approach requires to know the probability distribution of uncertain data, 
 which is often difficult to obtain.
Also, to ensure the solution quality, in stochastic programming, we need to generate a large set of scenarios based on this probability distribution and 
associate each scenario with a certain probability. Hence, even if the distribution is known, the stochastic model can still be  computationally prohibitive, and even intractable.

Recently, robust optimization (RO) \cite{RObook} has  emerged as an alternative methodology to handle data uncertainty. Since the RO approach does not require  knowledge of probability distribution of the uncertainty, it can avoid some of the difficulties arising from
the stochastic programming approach. Indeed, RO has also been applied in the context of cloud resource management  \cite{scha10,rkae13,yli18}. Unlike stochastic programming where uncertainty is captured by a large number of scenarios, uncertainty in RO is described by parametric sets, called uncertainty sets. 
Since  uncertainty sets can be constructed simply by using information such as lower bounds and upper bounds of uncertain parameters (i.e.,  random variables), 
it is much easier to derive than exact probability distributions.

The goal of RO is to find a robust solution that not only optimizes system performance but can also hedge against any perturbation in the input data within the uncertainty sets. Thus, the solution to a RO model tends to be conservative. 
However, the conservativeness of  robust solutions can be controlled by adjusting the uncertainty sets whose forms significantly affect the tractability and computational complexity of a robust model \cite{RObook}.
Furthermore, while a larger uncertainty set strengthens the robustness of a solution, it also increases the conservativeness. 
In practice, RO models usually scale well with the increasing dimension of data and are computationally tractable for large-scale systems. Additionally,  uncertainty sets are often constructed based on the desired level of robustness, historical data and experience of the decision maker.

In RO models, all decisions are made before the uncertainty is revealed, which can be overly conservative. 
To tackle this issue, adaptive robust optimization (ARO) \cite{ARO}, also known as two-stage RO \cite{2RO}, has recently been  introduced, where the second-stage problem models recourse decisions  after
observing the first-stage decisions and the realization of the uncertainty.
The first-stage decisions are often referred as  “\textit{here-and-now}”  decisions that cannot  be  adjusted  after  the uncertainty  is  disclosed,  while  the  second-stage decisions are known as ``\textit{wait-and-see}'' decisions that can be adjusted and adapted to the actual realization of  uncertain data.
Thus,  ARO  is still robust while less conservative than  RO.
 
In this paper, to address the challenge  caused by the demand uncertainty,  we propose a novel two-stage RO model to help the SP  identify optimal ENs for placing the service  and optimal amount of resource procured from each node before knowing the actual demand.
The service placement and sizing  are the first-stage decisions that 
are robust against any  realization of  traffic demand within a predefined   uncertainty set.
 Given the  first-stage  solution, the optimal  workload allocation decision is made in the second-stage after the uncertainty is disclosed.
 
 The rationale behind this design is that service placement and sizing typically happen at a larger time scale (e.g.,  in the order of hours or days) to ensure system stability \cite{vfar19} while the workload allocation decisions can  be adjusted in a shorter time scale (e.g., every few minutes) based on the actual demand. 
 Furthermore, in practice, the SP may not be able to change the 
 resource procurement decision frequently in short time scale.
 Hence, the two-stage RO approach is a  reasonable modeling choice for our problem. 
 Note that the first-stage decision  is robust against  all  scenarios, including the worst-case one, contained in  the uncertainty set. If the realized demand is not the worst case, the workload allocation can be updated based on the actual demand in the operation stage.

To the best of our knowledge, 
this is the first two-stage robust model for the edge service placement and sizing problem. 
Our main contributions are summarized below:

\begin{itemize}
\item We first introduce a deterministic MILP model for joint edge service placement, resource procurement, and workload allocation, which is then extended to a  new two-stage RO model to deal with demand uncertainty. In particular, the first-stage decision variables include  service placement and resource sizing, while resource allocation and request scheduling are the second-stage decisions.

\item The formulated model is a trilevel optimization problem that is decomposed into a master problem and a max-min subproblem. 
The bilevel subproblem is reformulated into a single-level problem with complementary constraints, which is then transformed into an MILP. 
We develop an iterative algorithm based on the column and constraint generation (CCG) method \cite{2RO} to solve the problem in a master-subproblem framework, which is guaranteed to converge in a finite, typically small, number of iterations.

\item Finally, extensive numerical  results are presented to demonstrate the efficacy of the proposed ARO model compared to the deterministic and RO models. We also perform sensitivity analysis to evaluate the impacts of different system  parameters on the optimal solution.
\end{itemize}

 The rest of the paper is organized as follows. In  Section \ref{model}, we describe the system model. 
In Section \ref{modfor}, we fist formulate the deterministic optimization model, which is then extended to a static robust model and an adaptive robust  model. The CCG-based iterative solution approach is introduced in Section \ref{sol}. Simulation results are shown in Section \ref{sim}
followed by discussion of related work in Section \ref{rel}.
Finally, we present conclusions and future research directions in Section \ref{conc}.

\section{System Model }
\label{model}

\begin{table}[ht] 
\centering
\caption{NOTATIONS}
\begin{tabular}{|l|l|}
\hline
Notation   & Meaning\\
\hline	
EN, AP & Edge Node, Access Point\\
\hline	
$\mathcal{M}$, M & Set and number of APs\\
\hline
$\mathcal{N}$, N & Set and  number of ENs\\
\hline	
$i$, $j$   & AP index and EN index\\
\hline	
$C_j$ & Resource capacity  of EN $j$ available for purchase\\
\hline
$p_0$ & Price of one computing unit at the cloud\\
\hline
$p_j$ & Price of one computing unit at EN $j$\\
\hline
$f_j$ & Service placement cost at  EN $j$ \\
\hline
$s_j$ & Storage cost at  EN $j$ \\
\hline
$D^{\sf m}$ &   Delay threshold  \\                  
\hline
$d_{i,0}$ & Network delay between AP $i$ and the cloud\\
\hline
$d_{i,j}$ & Network delay between AP $i$ and EN $j$\\
\hline
$B$ & Budget of the service provider \\
\hline
$\beta$ & Weighting factor, delay cost\\
\hline
$w$ & Computing resource demand of one request\\
\hline
$\lambda_i^{\sf f}$ & Forecast traffic demand at AP $i$\\
\hline
$\lambda_i$ & Actual traffic demand  at AP $i$\\
\hline
$\mathcal{C}^{\sf d}$ & Total network delay cost \\
\hline
$\mathcal{C}^{\sf c}$ & Cloud resource procurement cost \\
\hline
$\mathcal{C}_j^{\sf p}$ & Service placement  cost at EN $j$\\
\hline
$\mathcal{C}_j^{\sf e}$ & Edge  resource procurement cost at EN $j$\\
\hline
$\mathcal{C}_j^{\sf s}$ & Storage cost at EN $j$\\
\hline
$x_{i,0}$ & Workload at AP $i$ assigned to the cloud \\
\hline
$x_{i,j}$ & Workload  at AP $i$ assigned to EN $j$\\
\hline
$y_0$ & Amount of  computing resource purchased at the cloud\\
\hline
$y_j$ & Amount of  computing resource purchased at EN $j$\\
\hline
 $z_j^0$ & $\{0,1\}$, 1 if the service is placed at EN $j$ at the beginning \\
\hline
$z_j$ &  Binary variable, 1 if the service is placed at EN $j$  \\
\hline
$\Gamma$ & Uncertainty budget\\
\hline
$\mathcal{D}$ & Demand uncertainty set\\
\hline
\end{tabular} \label{notation}
\end{table}

In this paper, we study the service placement and sizing from the perspective of a SP (e.g., Google Maps, AR/VR). The SP  has
subscribers located in different areas, where  user requests in each area are aggregated at an access point (AP). Without EC, the requests are typically sent to remote servers for processing. However, with EC,
the SP can serve the requests at ENs 
closer to the users. We envision 
the emergence of an EC market  managed by a Telco, a cloud provider (e.g., Amazon), or a third-party \cite{duong1,duong2,duong3}. Indeed, numerous EC markets are currently being constructed by big companies and startups. The SP is assumed to have access to a portal listing various types of ENs in the market. Based on different factors such as price and location, the SP will decide where to place the service and how much resource to buy from each EN.

\begin{figure}[h!]
	\centering
		\includegraphics[width=0.40\textwidth,height=0.135\textheight]{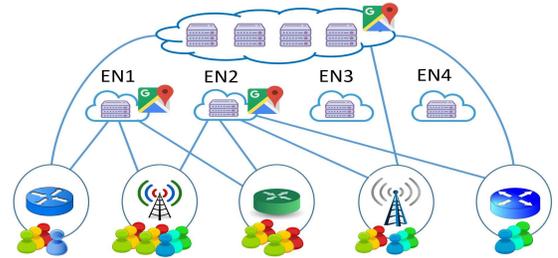}
			\caption{System model}
	\label{fig:model}
\end{figure}

We consider an EC system that consists of a set $\mathcal{N}$ of $N$ ENs and a set $\mathcal{M}$ of $M$ APs.  
Let $i$ and $j$ be the EN index and AP index, respectively. Also, denote by $C_j$ the computing resource capacity of   EN $j$ (e.g., number of servers, number of virtual machines, number of vCPUs, or number of CPU cycles per second). Define  $p_0$ and $p_j$ as the price of one computing unit at the cloud and at EN $j$, respectively.
In practice, the ENs can be geographically distributed in different locations that  
have different electricity prices. 
Furthermore, the ENs may come with different sizes, ownership, server types, and configurations.
 Thus, the computing resource prices can vary among the ENs. 
 Similar to the previous literature \cite{duong1,duong2,duong3,ryu19,lyan16,dzen16,spas19,qfan18,vfar19,lwan18,zxu16,mjia17,lma17, ayou19,bgao19,nkhe19,kpout19,touy19,lche18,lzha18,syan19,aces17,smon18,fhai19}, we  only consider the network from the APs to the ENs and the  cloud. 
 Let $d_{i,0}$ and $d_{i,j}$ be the distance between AP $i$ and the cloud, and between AP $i$ and AP $j$, respectively.
Fig.~\ref{fig:model} illustrates the system model with 4 ENs and 5 APs. The service here is Google Maps, which is placed at  EN1,  EN2, and the cloud.

The SP is assumed to be a price-taker and have a budget $B$ for operating the service at the edge and the cloud.
The SP uses the budget to buy cloud and  edge resources to serve the user requests during a fixed period of time (e.g., 30 minutes,  few hours, or a day), 
with the goal of minimizing not only the resource procurement cost but also  the total network delay of the  requests.
As mentioned in the introduction, to lower the network delay and reduce the data transmission, the SP can place the service on different ENs. 
Let $z_j$ be a binary  variable that  equals to $1$ if the SP decides to place the service onto EN $j$ and $0$ otherwise. Also, let $z_j^0$ be a binary indicator that  equals to $1$ if the service is available at EN $j$ at the beginning. 

If the SP wants to place the service onto  EN $j$  that does not have the service installed at the beginning of the scheduling horizon, we need to download the software and data of the service from a remote server or nearby ENs and install it onto EN $j$. This will incur a  cost $f_j$ that can be calculated as follows. First, denote by $\mathcal{F}$ the set of ENs that have the service at the beginning, 
i.e., $\mathcal{F} = \{j ~|~ z_j^0  = 1 \}$. Let $f_{j,0}$  
be the cost of installing and downloading the  service from the cloud  to EN $j$. Similarly,  $f_{j,j'}$ is the  cost of installing and downloading the  service from EN $j'$ to EN $j$. Typically, $f_{j,0} > f_{j,j'}$ for all $j$ and $j'$ because the distance between two ENs is much shorter than the distance between the cloud and an EN.
We have:
\beqn
f_j = \text{min} \Big\{f_{j,0}, \underset{j' | j' \in  \mathcal{F}}{\text{min}} f_{j,j'} \Big\}.
\eeqn

Let $s_j$ be the cost of storing the service at EN $j$. Also, define $y_0$ and $y_j$ as the amount of computing resource purchased from the cloud and EN $j$, respectively. The request arrival rate (i.e., traffic demand, workload) of the service  at AP $i$ is denoted by $\lambda_i$.  
Given the set of procured edge and cloud resources,
the SP decides how to optimally divide the workload in different areas to the ENs and the cloud for processing
to minimize the total network delay.
Define $x_{i,0}$ as the portion of workload  at AP $i$ to be routed to the cloud, and  $x_{i,j}$ as  the workload at AP $i$ assigned to EN $j$. Obviously, the SP  prefers to have the requests  processed by an EN closer to them rather than the remote cloud. To ensure the service quality, the SP may impose an upper bound $D^{\sf m}$ on the average network delay. A lower average delay implies that more requests are processed at the edge.
The main notations are summarized in Table \ref{notation}.

\section{Problem Formulation}
\label{modfor}

In this section, we first present a deterministic formulation of the  service placement,  resource procurement, and request scheduling problem, which is then extended to a two-stage robust formulation  to deal with demand uncertainty.

\subsection{Deterministic Formulation}

In the deterministic model, the SP jointly optimizes the service placement, sizing, and workload allocation decisions when the traffic demand  is assumed to be known exactly.
\subsubsection{Objective Function}

In this paper, we are interested in minimizing the total cost of the SP as follows:
\beqn
 \underset{x,~y,~z}{\text{minimize}} ~ \mathcal{C} =  \mathcal{C}^{\sf p} + \mathcal{C}^{\sf s}  + \mathcal{C}^{\sf e} + \mathcal{C}^{\sf c} + \mathcal{C}^{\sf d},
\eeqn
where $\mathcal{C}^{\sf p}$ is the placement  cost, $\mathcal{C}^{\sf s}$ is the storage cost, $\mathcal{C}^{\sf e}$ the  cost of purchasing edge computing resource, $\mathcal{C}^{\sf c}$ is the cloud resource procurement cost, and $\mathcal{C}^{\sf d}$ is the network delay cost. Also, we have $z = \big(z_1, z_2,\ldots, z_N \big)$, $y = \big(y_0, y_1,\ldots, y_N \big)$, $x = \big\{x_0, x_1, \ldots, x_N \big\}$, $x_0 = \big(x_{1,0}, x_{2,0},\ldots, x_{M,0} \big)$, and $x_j = \big(x_{1,j}, x_{2,j},\ldots, x_{M,j} \big),~\forall j$.


As explained in the system model section, the service placement cost $\mathcal{C}_j^{\sf p}$  is the total cost of downloading and installing the service at EN $j$, which can be expressed as: 
\beqn
\mathcal{C}_j^{\sf p} = f_j (1-z_j^0) z_j, \quad \forall j.
\eeqn
Clearly, if the service is  available at EN $j$ at the beginning (i.e., $z_j^0 = 1$), the service placement cost becomes zero. When the service is placed at EN $j$ but it is  not  available  at  the node at the beginning   (i.e., $z_j = 1$ and $z_j^0 = 0$), $\mathcal{C}_j^{\sf p}$ is equal to $f_j$. 
Additionally, if the SP decides to run the service at EN $j$ ($z_j = 1$), the cost of storing the service  data at EN $j$ is:
\beqn
\mathcal{C}_j^{\sf s} = s_j z_j, ~ \forall j.
\eeqn

The edge resource procurement cost at EN $j$ is equal to the amount of purchased resource $y_j$ multiplied by the resource price at the EN, i.e., we have:
\beqn
\mathcal{C}_j^{\sf e} = p_j y_j, ~ \forall j.
\eeqn
Similarly, the cloud resource procurement cost
$\mathcal{C}^{\sf c}$ is $p_0 y_0$. 
In addition, the total network delay of all the service requests is:
\beqn
d^{\sf tot} = \sum_i d_{i,0} x_{i,0} + \sum_{i, j}  d_{i,j} x_{i,j}.  
\eeqn
Thus, the total network delay cost $\mathcal{C}^{\sf d}$  is $\beta  d^{\sf tot} $. 
Overall, the goal of SP is to minimize the following objective function:
\beqn
\label{eq:obj}
\mathcal{C} = \sum_j f_j (1 - z_j^0) z_j + \sum_j s_j z_j + p_0  y_0  + \sum_j p_j y_j   \\ \nonumber
+~ \beta ~ \Big(\sum_i d_{i,0} x_{i,0} + \sum_{i, j} d_{i,j} x_{i,j}  \Big).
\eeqn

The SP can vary the delay cost parameter $\beta$  to control the tradeoff between the total expenses of the SP and the total network delay of the requests (i.e., between  cost and service quality).
The weight $\beta$ reflects the SP's attitude towards the network delay. Clearly, a larger $\beta$ indicates that the SP is more delay-sensitive and willing to spend more  to lower the delay. Note that the  delay of each request includes the transmission delay, propagation delay (network delay), and processing delay at the cloud or an EN. In this work, we assume that each request is assigned a fixed amount of computing resource $w$ and transmission bandwidth. Thus, the processing  delay  and the transmission delay (i.e., the request size divided by the bandwidth) are fixed. Hence, for simplicity, we consider the network delay only.  

 Additionally, it is straightforward to consider other costs such as bandwidth cost, which would discourage
 sending workload to the cloud, with minor modifications.
We are now ready to describe all the constraints of  the underlying optimization problem.

\subsubsection{Budget Constraint}
The total expense  should not exceed the budget of the SP.
 Thus, we have the following constraint:
\beqn
\label{eq:budget}
\sum_j f_j (1 - z_j^0)  z_j + \sum_j s_j z_j + \sum_j p_j y_j + p_0  y_0 \leq B.
\eeqn

\subsubsection{Reliability Constraint}  To enhance the service reliability, the SP may want to place the service on at least a minimum number of $r^{\sf min}$ ENs since link/node failures can occur unexpectedly. Hence, we can impose:
\beqn
\label{eq:reli}
\sum_j z_j \geq r^{\sf min}.
\eeqn

\subsubsection{Workload Allocation Constraints}
The service requests arriving at each AP must be allocated to either the remote cloud  or the ENs. Hence, we have:
\beqn
\label{eq:dwa}
x_{i,0}  + \sum_j x_{i,j} = \lambda_i, ~~\forall i.
\eeqn

\subsubsection{Capacity Constraints}

The  amount of computing resource $y_0$ and $y_j$ purchased at the cloud and each EN $j$ should be sufficient to serve the resource demand of all requests assigned to the cloud and the EN. 
Also, the   SP buys resource only from the ENs that have the service installed (i.e., $z_j = 1$). Furthermore, the amount of resource purchased from each EN $j$ cannot exceed the capacity $C_j$ of the EN. Thus:
\beqn
\label{eq:x0}
w \sum_i x_{i,0} \leq y_0 \\
\label{eq:xy}
w \sum_i x_{i,j} \leq y_j, \quad \forall j \\
\label{eq:capa}
 y_j \leq z_j C_j, \quad \forall j.
\eeqn

\subsubsection{Delay Constraint}
In order to ensure a certain service quality for their users, the SP may require the average network delay ($d^{\sf avg}$) of the requests does not exceed a certain maximum delay threshold $D^{\sf m}$. The average network delay is: 
\beqn
d^{\sf avg} = \frac{\sum_i d_{i,0} x_{i,0} + \sum_{i, j} d_{i,j} x_{i,j}  }{\sum_i \lambda_i}
\eeqn
Hence, $d^{\sf avg} \leq  D^{\sf m}$  can be rewritten as:
\beqn
\sum_i d_{i,0} x_{i,0} + \sum_{i, j} d_{i,j} x_{i,j}  
\leq~ D^{\sf m} \sum_i \lambda_i.
\eeqn

\subsubsection{Constraints on Variables}
The service placement indicator $z_j$ is a binary variable. Also, the workload allocation  $x$ and resource procurement $y$ must be non-negative. Thus: 
\beqn
x_{i,0} \geq 0,~\forall i;~~ x_{i,j} \geq 0,~\forall i,j \\ 
\label{eq:vars}
z_j \in \{0, 1\},~\forall j;~~y_0 \geq 0;~ y_j \geq 0,~\forall j.
\eeqn

Overall, the SP aims to solve the following problem:
\beqn
\textbf{(Deter)} &&\underset{x,~y,~z}{\text{minimize}} ~~ \mathcal{C}  \nonumber \\ \nonumber
&& \text{subject to} ~~ (\ref{eq:obj})-(\ref{eq:vars}).
\eeqn

This is an MILP that can be solved efficiently using existing MILP solvers.
Note that our model can be easily extended to capture other  system and design constraints. For instance, we may consider multiple resource types (e.g., RAM, CPU, bandwidth) instead of only computing resource. The SP may be enforced to buy an integer quantity of computing units from each EN rather than a continuous amount $y_j$, which can be easily handled using a unary  or binary expansion.

In summary, given the system parameters such as the network topology, the downloading and installation costs, the resource price and capacity at each EN,  the budget, and the delay penalty cost,  the SP solves the MILP problem (Deter) to find an optimal service placement, resource procurement, and workload allocation solution.

\subsection{Uncertainty Modeling}
In the deterministic model,
the traffic demand is assumed to be known exactly at the time 
of making the service placement and sizing decision. 
In other words, the SP assumes that the actual demand is the same as the forecast one, which  is then used as input to the deterministic problem.
However, the exact demand in each area 
typically cannot be accurately predicted at the time of making the strategic decision. Thus, how to properly capture the uncertainties in the decision making process is a crucial task.

In RO, uncertain parameters are modeled through uncertainty sets, which express an infinite number of scenarios.
 A well constructed uncertainty set should be computationally tractable and balance robustness and conservativeness of the robust solution \cite{RObook}.
 In practice, the polyhedral uncertainty set is a natural and popular choice for representing uncertainties in the RO literature \cite{RObook,ARO,2RO}. In particular, to construct a polyhedral uncertainty set, the SP would need to specify  intervals expressing the uncertain demand at every AP, and a parameter to control the degree of conservativeness.

Define $\lambda = (\lambda_1, \lambda_2,\ldots, \lambda_M)$ and 
$\lambda^{\sf f} = (\lambda_1^{\sf f}, \lambda_2^{\sf f},\ldots, \lambda_M^{\sf f})$ as  the actual demand vector  and the forecast demand vector, respectively. Note that $\lambda_i^{\sf f}$ is also called the nominal value of the uncertain demand $\lambda_i$. Additionally, let $\hat{\lambda} = (\hat{\lambda}_1, \hat{\lambda}_2,\ldots, \hat{\lambda}_M)$, where $\hat{\lambda}_i$ is the maximum demand deviation, which can be understood as the maximum forecasting error of the uncertain demand at AP $i,~\forall i$.
Thus, $[\lambda_i^{\sf f} - \hat{\lambda}_i, \lambda_i^{\sf f} + \hat{\lambda}_i]$ represents the uncertain demand at AP $i$. 
The polyhedral  uncertainty set can be defined as follows:
\beqn
\mathcal{D}(\lambda^{\sf f}, \hat{\lambda}, \Gamma) = \Big\{   
&&\lambda_i = \lambda_i^{\sf f} + g_i \bar{\lambda}_i,~\forall i; ~g_i \in [-1,~1],~\forall i \nonumber\\ 
&& \sum_i~ | g_i | \leq \Gamma  \quad
\Big\},
\label{uncertainty_budget}
\eeqn
where $\Gamma$ is the budget of uncertainty,
which can vary in the continuous interval [0, M]. The form of this uncertainty set  is widely used in the RO literature \cite{RObook,ARO,2RO}.
This set contains the lower and upper bounds of the uncertain parameters and the bound
of the linear combination of the uncertain parameters. These
information can be extracted by learning from historical data.
 Indeed, a more general polyhedron can also be accommodated by the RO approach. 
 
 The actual demand $\lambda_i$ can take any value in the range of 
 $[\lambda_i^{\sf f} - \hat{\lambda}_i, \lambda_i^{\sf f} + \hat{\lambda}_i]$. However, in practice, it is not likely that all the actual demands are simultaneously close to the corresponding lower bounds or upper bounds. This observation is captured by the 
 uncertainty budget $\Gamma$.
A larger value of $\Gamma$  implies a larger uncertainty set and a more robust solution. However, the solution is also more conservative to  protect the system against a higher degree of uncertainty. Therefore, $\Gamma$ can be used to adjust the  robustness against the conservative level of the solution. 
When $\Gamma = 0$, 
the actual demand is equal to the forecast demand and a robust model becomes a deterministic model without considering demand uncertainty.
When $\Gamma= \mbox{M}$, we simply consider all possible realizations of the uncertain demand $\lambda_i$ in the interval of $[\lambda_i^{\sf f} - \hat{\lambda}_i, \lambda_i^{\sf f} + \hat{\lambda}_i]$, and 
$\mathcal{D}$ becomes a box uncertainty set. Hence, the polyhedral uncertainty set is less
conservative compared to the box uncertainty set.

\subsection{Robust Optimization Formulation}

In the static (single-stage) robust optimization model, the service placement,  sizing, and workload allocation decisions are made simultaneously before observing the actual realization of the demand. Based on the RO theory \cite{RObook}, the robust service placement and sizing problem can be formulated as:
\beqn
\label{ROobj}
 \underset{(x,y,z) \in  \mathcal{S}_{(X,Y,Z)}  }{\text{min}} ~\underset{\lambda \in \mathcal{D}}{\text{max}} ~ \sum_j f_j (1 - z_j^0) z_j + \sum_j s_j z_j +  p_0  y_0    \nonumber \\ 
   +  \sum_j p_j y_j + \beta  \Bigg\{ \sum_i d_{i,0} x_{i,0} + \sum_{i,j}   d_{i,j} x_{i,j} \Bigg\} 
\eeqn
subject to 
\beqn 
\label{demandRO}
x_{i,0}  + \sum_j x_{i,j} \geq \lambda_i, \quad \forall i, \lambda \in \mathcal{D}    \\ 
\label{delayRO}
\sum_i d_{i,0} x_{i,0} + \sum_{i,j} d_{i,j} x_{i,j}   \leq D^{\sf m} \sum_i \lambda_i, \lambda \in \mathcal{D}, 
\eeqn
where $\mathcal{S}_{(X,Y,Z)}$ is the set of constraints on the placement, sizing, and workload allocation variables. These constraints are given in the deterministic model. The RO model aims to minimize the total cost of the SP under the worst-case demand scenario. Also, all the constraints related to  uncertain parameters should be satisfied for any potential realization of these parameters  within the uncertainty set. 

Note  that  in  the  deterministic  formulation, 
the workload allocation constraints (\ref{eq:dwa}) can be written as either equalities or  inequalities
since inequalities become equalities at the optimality for the cost minimization objective.
Also, equalities related to uncertainties are meaningless in the RO approach \cite{RObook} since the equality constraints cannot be satisfied for every demand scenario in the uncertainty set. Hence, the workload allocation constraints is written in form of inequalities as in (\ref{demandRO}).
Due to space limitation, we do not present the solution approach here and refer to \textit{Appendix} \ref{ROmodelA} for more details.

\begin{figure}[h!]
	\centering
		\includegraphics[width=0.44\textwidth,height=0.18\textheight]{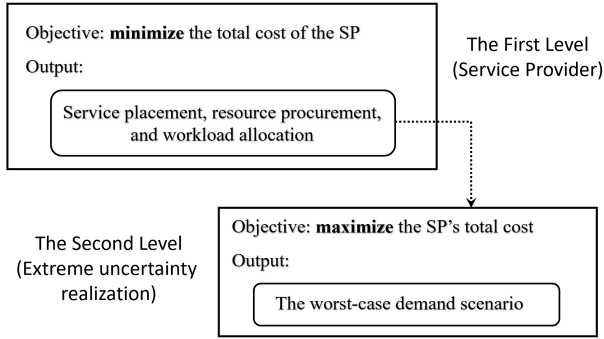}
			\caption{The RO model for edge service placement and sizing}
	\label{fig:romodel}
\end{figure}

Overall, the static single-stage robust model  includes two levels, as  illustrated in Fig. \ref{fig:romodel}. The first level represents the decision-making problem of the SP prior to the uncertainty realization and seeks to minimize objective function value (i.e., the total cost of the SP). The second level represents the uncertainty realization in the worst case within the  uncertainty set $\mathcal{D}$ and aims to maximize the objective function value. 
 Given the uncertainty set and the system parameters as the input to the decision-making problem, the SP solves the RO model (\ref{ROobj})-(\ref{delayRO}) to determine a robust optimal service placement and sizing solution.

 \vspace{-0.1in}

\subsection{Two-Stage Adaptive Robust Formulation}

Unlike  the RO model where all decisions are made simultaneously in a single stage, the ARO model includes two stages. In particular, the service placement and the resource sizing  are the first-stage decision variables. 
Given the first-stage decisions, the workload allocation decisions (i.e.,  operation decisions) are determined as an optimal solution to the second-stage problem (i.e., the recourse problem). Note that the first-stage decisions are made without the knowing the uncertain parameters while the recourse decisions are made based on the revealed uncertainties. 
Define $h_j = f_j (1 - z_j^0) + s_j,~\forall j.$ The two-stage adaptive robust  model can be formulated as:

\beqn
\label{AROori}
&&\underset{(y,z) \in \mathcal{S}_{(Y,Z)}}{\text{min}}~~ \Bigg\{~~ \sum_j h_j z_j + \sum_j p_j y_j + p_0  y_0 ~~~~ \quad  ~~~~  \\ \nonumber
&& ~~~~\quad + \underset{\lambda \in \mathcal{D}}{\text{max}} ~  \underset{x \in \mathcal{F}(y,\lambda)}{\text{min}} 
~\beta  \Big( \sum_i d_{i,0} x_{i,0} + \sum_{i,j} d_{i,j} x_{i,j} \Big)  \Bigg\},
\eeqn
where:
\beqn
\mathcal{S}_{(Y,Z)} = \Bigg\{  
&&\sum_j z_j \geq r^{\sf min};~~y_j \leq z_j C_j, \quad \forall j, \\ \nonumber 
&&\sum_j h_j  z_j  + \sum_j p_j y_j + p_0  y_0  \leq B, \\ \nonumber
&&z_j \in \{0,~1\};~~y_0 \geq 0;~~y_j \geq 0,~~\forall j  \nonumber
\Bigg\} 
\eeqn

\beqn
\label{eq:Fylambda}
\mathcal{F}(y,\lambda) = \Bigg\{
&& x \in \mathcal{S}_X;~~x_{i,0}  + \sum_j x_{i,j} = \lambda_i, ~ \forall i, \\ \nonumber
&&w \sum_i x_{i,0} \leq y_0;~~ w \sum_i x_{i,j} \leq y_j, ~ \forall j, \\ \nonumber
&&\sum_i d_{i,0} x_{i,0} + \sum_{i,j} d_{i,j} x_{i,j}  \leq D^{\sf m} \sum_i \lambda_i 
\Bigg\}
\eeqn
\beqn
\label{eq:SX}
\mathcal{S}_X  = \Big\{ 
&&x_{i,0} \geq 0,~~\forall i;~~x_{i,j} \geq 0,~~ \forall i,j~~ \Big\}.
\eeqn

Note that $\mathcal{S}_{(Y,Z)}$ represents the constraints on the service placement  and sizing variables, and $\mathcal{F}(y,\lambda)$ is the set of feasible workload allocation solutions for a fixed 
resource procurement decision $y$ and demand realization $\lambda$. The optimal workload allocation aims to minimize the total delay cost in the worst-case scenario of the demands.
The two-stage robust service placement and sizing above is inherently a trilevel (mix-max-min) optimization model, which is difficult to solve.

\begin{figure}[h!]
	\centering
		\includegraphics[width=0.45\textwidth,height=0.22\textheight]{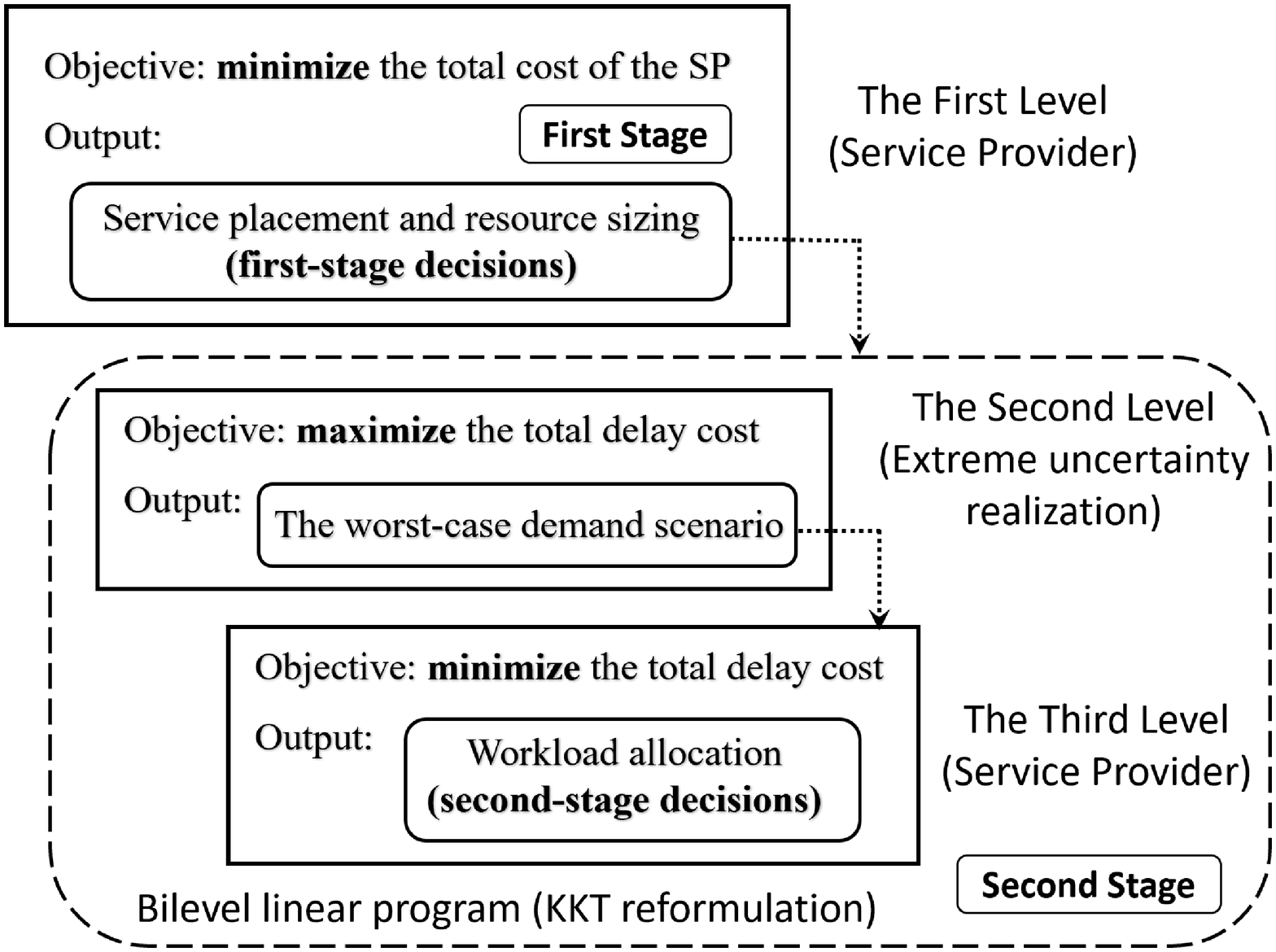}
			\caption{The proposed tri-level ARO model }
	\label{fig:aromodel}
\end{figure}

Fig. \ref{fig:aromodel} illustrates the  ARO model. The first level represents the decision-making problem of the SP before the uncertainty realization and seeks to minimize the SP's  total cost. The second level represents the worst-case uncertainty realization, 
 which tries to degrade the service quality by maximizing the total network delay.  Finally, the third level represents the workload allocation decisions to mitigate the effects of the realized uncertainty. The workload allocation problem  aims to minimize the total delay cost.
 \vspace{-0.05in}
  \vspace{-0.15in}

\section{Solution Approach}
\label{sol}

In this section, we develop an iterative algorithm based on the column and constraint generation (CCG) procedure \cite{2RO} 
to solve the formulated two-stage robust service placement and sizing problem (\ref{AROori}). In particular, the developed algorithm is implemented in a master-subproblem framework that decomposes the problem  into 
a master problem and a bilevel max-min subproblem. 
The optimal value of the master problem in each iteration provides a lower bound  while the optimal solution to the subproblem helps us compute an upper bound of the original two-stage robust problem (\ref{AROori}). The algorithm iteratively solves an updated master problem and an updated subproblem in each iteration until convergence.

First, it can be shown that the two-stage robust model (\ref{AROori}) can be transformed into an equivalent mixed integer program built on the collection of extreme points of the uncertainty set $\mathcal{D}$. It can be explained intuitively as follows. Assume that the problem (\ref{AROori}) as well as the innermost minimization problem in (\ref{AROori}) are feasible. Then, we can rewrite the second-stage bilevel problem as a max-max problem (i.e., simply a maximization problem) over 
$\lambda$ and the set  of dual variables associated with the constraints of the innermost minimization problem. 
Since this maximization problem is optimized over two disjoint polyhedrons, it always has an
optimal solution combining extreme points of these two polyhedrons \cite{EC2RO}. Therefore, the optimal solution to the second-stage problem always occurs an extreme point of the polyhedron  $\mathcal{D}$.

Define $K$ as the number of extreme points of  the uncertainty set $\mathcal{D}$. Also, let $\mathbb{D} = \{\lambda^1, \lambda^2, \ldots, \lambda^K \}$ be the set of extreme points (i.e., extreme demand scenarios)  of  $\mathcal{D}$, where $\lambda^l = \{\lambda_1^l, \lambda_2^l, \ldots, \lambda_M^l \}$ is the $l$-th extreme point. 
Therefore, the ARO model (\ref{AROori}) is equivalent to: 

\beqn
\label{AROex}
&&\underset{(y,z) \in \mathcal{S}_{(Y,Z)}}{\text{min}}~~ \Bigg\{~~ \sum_j h_j z_j + \sum_j p_j y_j + p_0  y_0 ~~~~~~~~~~~~~  \\ \nonumber
&& ~~~~\quad +~ \underset{\lambda \in \mathbb{D}}{\text{max}} ~  \underset{x \in \mathcal{F}(y,\lambda)}{\text{min}} 
~\beta  \Big( \sum_i d_{i,0} x_{i,0} + \sum_{i,j} d_{i,j} x_{i,j} \Big)  \Bigg\}.
\eeqn
Clearly, this problem can be transformed into the following equivalent MILP by enumerating all the extreme points in $\mathbb{D}$:
\beqn   
\label{objaro}
\underset{\eta, (y,z) \in \mathcal{S}_{(Y,Z)} }{\text{min}} ~~\sum_j h_j z_j + \sum_j p_j y_j + p_0  y_0 + \eta 
\eeqn
subject to
\beqn
\label{aroc1}
\eta \geq \beta  \Big\{ \sum_i d_{i,0} x_{i,0}^l + \sum_{i,j} d_{i,j} x_{i,j}^l \Big\}, ~~\forall l \leq K \\
\label{aroc2}
x^l \in \mathcal{F}(y,\lambda^l),~~ \forall l \leq K. 
\eeqn

For a large polyhedral uncertainty set as in (\ref{uncertainty_budget}), obtaining the optimal solution to the reformulated large-scale MILP above, which needs to enumerates all the possible extreme demand scenarios, may be not practically feasible.  
This motivates the iterative solution based on the CCG method  \cite{2RO}. Specifically, instead of solving the full problem (\ref{objaro})-(\ref{aroc2})
for all extreme points in the uncertainty set, we only solve this problem for a subset of $\mathbb{D}$, which obviously provides a valid relaxation of this problem and give us a  lower bound (LB).
Therefore, we can obtain stronger LBs by gradually adding non-trivial demand scenarios to the relaxed problem.
This is indeed the core idea behind the CCG method, which expands a subset of $\mathbb{D}$ gradually and add an additional variable $x^k$ in each iteration $k$. 
Furthermore, an optimal solution to the second-stage problem for a fixed  ($y$, $z$)
 clearly provides an upper bound (UB) of the  two-stage robust problem (\ref{AROori}).
 
 In the following, we first describe the master problem that is a relaxation of the problem (\ref{objaro})-(\ref{aroc2}). Then, we elaborate how to solve the subproblem (i.e., the second-stage problem) given the first-stage decision. Finally, we present the iterative algorithm for solving the two-stage robust service placement and sizing problem in a master-subproblem framework.

\subsection{Master Problem}
The master problem \textbf{(MP)} at iteration $k$ is given  as: 
\beqn
\underset{y,z, \eta}{\text{min}} ~\sum_j h_j z_j + \sum_j p_j y_j + p_0  y_0 + \eta 
\eeqn
subject to
\beqn
\eta \geq \beta  \Big\{ \sum_i d_{i,0} x_{i,0}^l + \sum_{i,j}  d_{i,j} x_{i,j}^l \Big\}, \forall l \leq k \\
x_{i,0}^l  + \sum_j x_{i,j}^l = \lambda_i^{*,l}, \forall l \leq k \\ 
w \sum_i x_{i,0}^l \leq y_0,~\forall l \leq k;~~
 w \sum_i x_{i,j}^l \leq y_j,  \forall j, \forall l \leq k \\ 
\sum_i d_{i,0} x_{i,0}^l + \sum_{i,j} d_{i,j} x_{i,j}^l  \leq D^{\sf m} \sum_i \lambda_i^{*,l}, \forall l \leq k \\
x_{i,0}^l \geq 0,~ x_{i,j}^l \geq 0, \forall l \leq k;~
 (y,z) \in \mathcal{S}_{(Y,Z)}; ~~ \eta \in \mathbb{R},
\eeqn
where $\{ \lambda^{*,1},\lambda^{*,2},
\ldots,\lambda^{*,k} \}$ is the set of optimal solutions to the subproblem in all  previous iterations up to iteration $k$. Also, $\lambda^{*,l} = \{ \lambda_1^{*,l}, \lambda_2^{*,l}, \ldots, \lambda_M^{*,l}\}, \forall l$.
The optimal solution to this master problem includes the optimal placement ($z_j^{*,k+1}, \forall j$), sizing ($y_0^{*,k+1}$, $y_j^{*,k+1}, \forall j$),  delay cost ($\eta^{*,k+1}$), and workload allocation $ x^{*,l}, \forall l \leq k$.
Then, the optimal sizing decisions $y_0^{*,k+1}$ and $y_j^{*,k+1}, \forall j$ will serve as input to the subproblem in Section 
\ref{subrefor}.
Indeed, the master problem at iteration $k$ corresponds to $k$ extreme points of the uncertainty set $\mathcal{D}$.
Therefore, because each \textbf{MP} contains only
a subset of constraints of the  original two-stage RO formulation  (\ref{objaro})-(\ref{aroc2}), the optimal solution to an \textbf{MP} provides a LB of the original problem. 
We also achieve a stronger LB in every iteration since each new iteration adds more constraints to the \textbf{MP}. Thus:
\beqn
\label{eq:LBupdate}
LB = \sum_j h_j z_j^{*,k+1}  
+ \sum_j p_j y_j^{*,k+1} + p_0  y_0^{*,k+1} + \eta^{*,k+1}.
\eeqn

\subsection{Reformulation of the Subproblem}
\label{subrefor}

The max-min subproblem is  indeed a bilevel optimization problem, which is difficult to solve. To this end, we show how to
 reformulate the subproblem as a MILP problem that can be solved globaly using MILP solvers. 
Specifically, given $y$, the subproblem \textbf{(SP)} is:
\beqn
\label{SP}
\mathcal{Q}(y) =  \underset{\lambda \in \mathcal{D}}{\text{max}} ~ \underset{x \in \mathcal{F}(y,\lambda)}{\text{min}} 
\beta \Bigg\{  \sum_i d_{i,0} x_{i,0} + \sum_{i,j} d_{i,j} x_{i,j} \Bigg\}
\eeqn
From (\ref{eq:Fylambda}), the inner minimization problem can be written as: 
\beqn
\label{eq:SPIobj}
\underset{x}{\text{min}} 
~\beta  \Bigg\{ \sum_i d_{i,0} x_{i,0} + \sum_{i,j} d_{i,j} x_{i,j} \Bigg\}
\eeqn
subject to
\beqn
\label{eq:SPcs}
  w \sum_i x_{i,0} \leq y_0, \quad (\pi_0)\\
  w\sum_i x_{i,j} \leq  y_j, ~ \forall j \quad (\pi_j) \\ 
\sum_i d_{i,0} x_{i,0} + \sum_{i,j} d_{i,j} x_{i,j}  \leq  D^{\sf m} \sum_i \lambda_i,~(\mu) \\
 x_{i,0}  + \sum_j x_{i,j} = \lambda_i, ~~ \forall i, \quad (\sigma_i)\\ 
 x_{i,0} \geq 0,~~\forall i \quad (\xi_{i,0})\\ 
\label{eq:SPce}
 x_{i,j} \geq 0,~~ \forall i,j, \quad (\xi_{i,j})
\eeqn
where $\pi_0,\pi_j,\mu,\sigma_i,\xi_{i,0},\xi_{i,j}$ are the dual variables associated with constraints (\ref{eq:SPcs})-(\ref{eq:SPce}), respectively. 
Also,  $y$ are the optimal sizing solution to the latest \textbf{MP}, i.e., at iteration $k$, we have  $y_0 = y_0^{*,k+1}$ and $y_j = y_j^{*,k+1}, \forall j$.

Based on  the Karush–Kuhn–Tucker (KKT) conditions \cite{boyd}, we can infer that given $(y, \lambda)$,
the optimal solution $x$ to the innermost 
problem (\ref{eq:SPIobj})-(\ref{eq:SPce}) is any of the feasible solutions to the set of constraints (\ref{eqe:comps})-(\ref{eqe:compe1}).
Please refer to \textit{Appendix \ref{SPKKTA}} for more details.
Hence,  the \textbf{SP} (\ref{SP}) is equivalent to the following problem with complementary constraints:
\beqn
\label{eq:SPobjf}
\underset{x,\lambda,\pi_0,\pi_j,\mu,\sigma_i}{\text{max}}
~\beta  \Bigg\{ \sum_i d_{i,0} x_{i,0} + \sum_{i,j} d_{i,j} x_{i,j} \Bigg\}
\eeqn
subject to
\beqn
\label{eqe:comps}
   0 \leq \beta d_{i,0} + w \pi_0 + \mu d_{i,0} - \sigma_i ~\bot~ x_{i,0} \geq 0,~\forall i\\
 0 \leq  \beta d_{i,j} + w\pi_j + \mu d_{i,j} - \sigma_i  ~\bot~ x_{i,j} \geq 0,~\forall i,j \\
 0 \leq \big(y_0 - w \sum_i x_{i,0} \big) ~\bot~ \pi_0 \geq 0\\ 
 0 \leq \big(y_j - w \sum_i x_{i,j}\big) ~\bot ~\pi_j \geq 0, ~ \forall j  \\ 
 \label{eqe:compe}
 0 \leq   \big(  D^{\sf m} \sum_i \lambda_i - \sum_i d_{i,0} x_{i,0} - \sum_{i,j} d_{i,j} x_{i,j} \big) \\ \nonumber  \bot ~ \mu \geq 0 \\
  \label{eqe:compe1}
x_{i,0}  + \sum_j x_{i,j}  = \lambda_i,~~\forall i\\
\lambda_i = \lambda_i^{\sf f} + g_i \hat{\lambda}_i,~\forall i;~~\sum_i t_i \leq \Gamma;~~  t_i \leq 1,~~ \forall i \\
\label{eq:SPlasteq}
-t_i + g_i \leq 0,~~ \forall i;~~t_i + g_i \geq 0,~~ \forall i,  
\eeqn
where the last two constraints represent the  uncertainty set $\mathcal{D}$. 
Note that a complimentary constraint $0 \leq x \bot \pi \geq 0$ means $x \geq 0, \pi \geq$ and $x . \pi = 0$. Thus, it is a nonlinear constraint. Fortunately, this nonlinear complimentary constraint can be transformed into equivalent exact linear constraints by using the Fortuny-Amat transformation \cite{bigM}. Specifically, the complementarity  condition $0 \leq x \bot \pi \geq 0$ is  equivalent  to the following set of mixed-integer linear constraints:
\beqn
x \geq 0;~~ x \leq (1-u)M \\ \nonumber
\pi \geq 0;~~ \pi \leq  uM,~~ u\in\{0;1\},
\eeqn
where M is a sufficiently large constant.
By applying this transformation to all the complementary constraints (\ref{eqe:comps})-(\ref{eqe:compe}), we obtain an MILP that is equivalent to the subproblem (\ref{SP}).
The explicit form of this  MILP is given in \textit{Appendix \ref{MILPSP}}. Thus, the subproblem  can be solved using an MILP solver.

Denote by ($\lambda^{*,k+1}$,  $x^{*,k+1}$)  the optimal solution to the \textbf{SP} at iteration $k$. 
The solution to each \textbf{SP}  helps us determine an UB to the original two-stage RO problem.
Specifically, we have:
\beqn
\label{eq:UBupdate}
UB = \min \Big\{UB,~UB^{k+1} \Big\}, \\
\label{eq:UBupdate1}
UB^{k+1} =  \sum_j h_j z_j^{*,k+1}   + \sum_j p_j y_j^{*,k+1} + p_0  y_0^{*,k+1}  \\ \nonumber
+ \beta  \Big( \sum_i d_{i,0} x_{i,0}^{*,k+1}  
+ \sum_{i,j} d_{i,j} x_{i,j}^{*,k+1} \Big).
\eeqn
Also, $\lambda^{*,k+1}$ is used as input to the \textbf{MP} in the next iteration.

\subsection{Algorithm}
Based on the description of the master problem and the subproblem, we are now ready to present the CCG-based iterative algorithm for solving the formulated two-stage robust service placement and sizing problem (\ref{AROori}) as shown in \textbf{Algorithm}
\ref{AROalg}.

\begin{algorithm}[ht!]
\caption{\textsc{ Two-Stage Adaptive Robust Algorithm}}
\label{AROalg}
\begin{algorithmic}[1]
\STATE Initialization:  set  $k=1$, $LB = -\infty$, and $UB =+\infty$. 
\REPEAT 
  \STATE Solve the following MP.
\beqn   
\label{algobjaro}
\underset{\eta, (y,z) \in \mathcal{S}_{(Y,Z)} }{\text{min}} ~~\sum_j h_j z_j + \sum_j p_j y_j + p_0  y_0 + \eta  \\
\label{algaroc2}
\text{subject to}~~~~x^l \in \mathcal{F}(y,\lambda^l),~~ \forall l \leq k ~~~~~~~~~~~\nonumber
\eeqn
\beqn
\label{algaroc1}
\eta \geq \beta  \Big\{ \sum_i d_{i,0} x_{i,0}^l + \sum_{i,j} d_{i,j} x_{i,j}^l \Big\}, ~\forall l \leq k \nonumber.  
\eeqn
Obtain an optimal solution $\big( \eta^{*,k+1},z^{*,k+1},y^{*,k+1},$\\$x^{*,1},\ldots, x^{*,k} \big)$  and update LB according to (\ref{eq:LBupdate}).
  \STATE Solve SP (\ref{SP}) with $y = y^{*,k+1}$ to obtain the worst-case demand $\lambda^{*,k+1}$ given $y$ and update UB following (\ref{eq:UBupdate}).

  \STATE Update $\lambda^{k+1} = \lambda^{*,k+1}$, and $k := k+1$. 
\UNTIL {$\frac{UB -LB}{UB} \leq \epsilon$ }\\
\STATE 
Output: optimal placement and sizing decisions $(z^*, y^*)$.
\end{algorithmic}
\end{algorithm}

Different from \cite{2RO}, we consider the extreme scenario with the maximum total demand in the first iteration. In particular, without loss of generality, let $\lambda^{\sf 1}$ be the extreme demand scenario  in $\mathcal{D}$ with the maximum total demand. Specifically, $\lambda^{\sf 1} = \{\lambda_1^{\sf 1}, \ldots, \lambda_M^{\sf 1} \}$ can be computed as an optimal solution to the following optimization problem:
\beqn
\label{maxDemand}
&&\underset{g,\lambda}{\text{max}}~ \sum_i \lambda_i \nonumber \\
&&\textit{s.t.}~~\lambda_i = \lambda_i^{\sf f} + g_i \hat{\lambda}_i,\forall i;
  -1 \leq g_i \leq 1,\forall i ;~\sum_i | g_i | \leq \Gamma.  \nonumber
\eeqn
Indeed, $\lambda^{\sf 1}$ can be found analytically by sorting the demand deviations $\hat{\lambda}_i$ in  descending order and set $g_i$ associated with the  $\floor{\Gamma}$ largest demand deviations to be 1 and $g_i$ of the next largest $\hat{\lambda}_i$ to be $\Gamma - \floor{\Gamma}$.
Note that the ARO model (\ref{AROori}) is equivalent to the problem
(\ref{objaro})-(\ref{aroc2}) whose constraints enumerate over all extreme points of $\mathcal{D}$, including $\lambda^{\sf 1}$. By considering $\lambda^{\sf 1}$ in the first iteration, we have $y_0 + \sum_j y_j \geq w \sum_i \big( x_{i,0}^{\sf 1} + \sum_j x_{i,j}^{\sf 1} \big) = w \sum_i \lambda_i^{\sf 1}$. Hence, $y_0 + \sum_j y_j
\geq w \sum_i  \lambda_i^l,\forall l \leq K$. As a result, 
the total resource purchased from the cloud and the ENs
is always sufficient to serve every realization of the  demand.

Since new constraints related to $\lambda^k$ are added to the MP (\ref{algobjaro}) at every iteration $k$, the LB is improved (weakly-increasing) at every iteration. Also, by definition as in (\ref{eq:UBupdate}),  the UB is  non-increasing.
Furthermore, as explained before, the worst-case demand in Step 4 is always an extreme point of the polyhedral uncertainty set $\mathcal{D}$. The set  $\mathbb{D}$ of extreme points is a finite set with $K$ elements. Hence, we can prove that \textbf{Algorithm}
\ref{AROalg} converges to the optimal value of the original two-stage robust problem (\ref{AROori}) in $O(K)$ iterations (see \textit{Appendix} \ref{proofcon}).

\section{Numerical Results}
\label{sim}

\subsection{Simulation Setting}
\label{setting}
Since EC is still in its early stage and we are not aware of any public data for  edge network topologies, similar to  previous works \cite{mjia17,mjia18}, we adopt the  Barabasi-Albert model \cite{abnet} to  
generate random a scale-free edge network topology 
with 100 nodes and the attachment rate of 2 \cite{mjia17}.
Also, link delays are randomly generated in between 2 ms and 5 ms \cite{zxu18}. 
The network transmission delay between any two APs is the delay of the shortest path between them. 
Based on the generated topology, 80 nodes are chosen as APs and 20 nodes are chosen as ENs. The delay between each AP and the remote cloud is set to be 80 ms. 
Additionally, the forecast traffic arrival rate (i.e., demand) at each AP is randomly drawn from 1000 to 4000 requests per time unit \cite{zxu18},
 and the  resource demand $w$ of each service request is set to be 1 MHz \cite{ratedata}.

Each EN is chosen randomly from the set of  Amazon EC2 M5 instances \cite{EC2}. Using the hourly
price of a general purpose m5d.xlarge Amazon EC2 instance  \cite{EC2} as a reference, the resource price (\$ per vCPU per time unit) at the cloud is set to be 0.03 while the resource prices at the ENs are randomly generated from 0.04 to 0.06.  Additionally, the service placement costs ($f_j$) and storage costs ($s_j$) at the ENs are randomly generated in the ranges of [0.2, 0.25] and [0.1, 0.12], respectively. The budget of the SP is set to be 100. We also assume that the service is not available on any EN at the beginning (i.e., $z_j^0 = 0, \forall j)$. 

For the sake of clarity in the figures and analysis, in the \textbf{base case}, we consider a small system with 20 APs and 5 ENs (i.e., M = 20, N = 5), which are selected randomly from the corresponding original sets of 80 APs and 20 ENs.
Note that we  also study the impacts of varying the number of APs and ENs later. In the base case,   
the maximum average delay $D^{\sf m}$ is set to be 30 ms, and the minimum number of edge servers $r^{\sf min}$ is  2. Let $\alpha_i$ be the ratio between the maximum demand deviation $\hat{\lambda}_i$ and the forecast demand $\lambda_i^{\sf f}$.  Also, define $\beta_0 = \beta \sum_i \lambda_i^{\sf f}$. In the base case, we set  $\alpha_i = \alpha = 0.3,~\forall i$, the uncertainty budget $\Gamma = 10$, and the delay penalty parameter $\beta_0 = 0.01$.
This default setting 
is used in most of the
simulations unless mentioned otherwise. We implement all the algorithms in Matlab environment using CVX\footnote{http://cvxr.com/cvx/} and Gurobi\footnote{https://www.gurobi.com/}.

\subsection{Performance Evaluation}

\subsubsection{Comparison between  RO, ARO, and deterministic models}

\begin{figure}[h!]
		\subfigure[Cost]{
		  \includegraphics[width=0.245\textwidth,height=0.12\textheight]{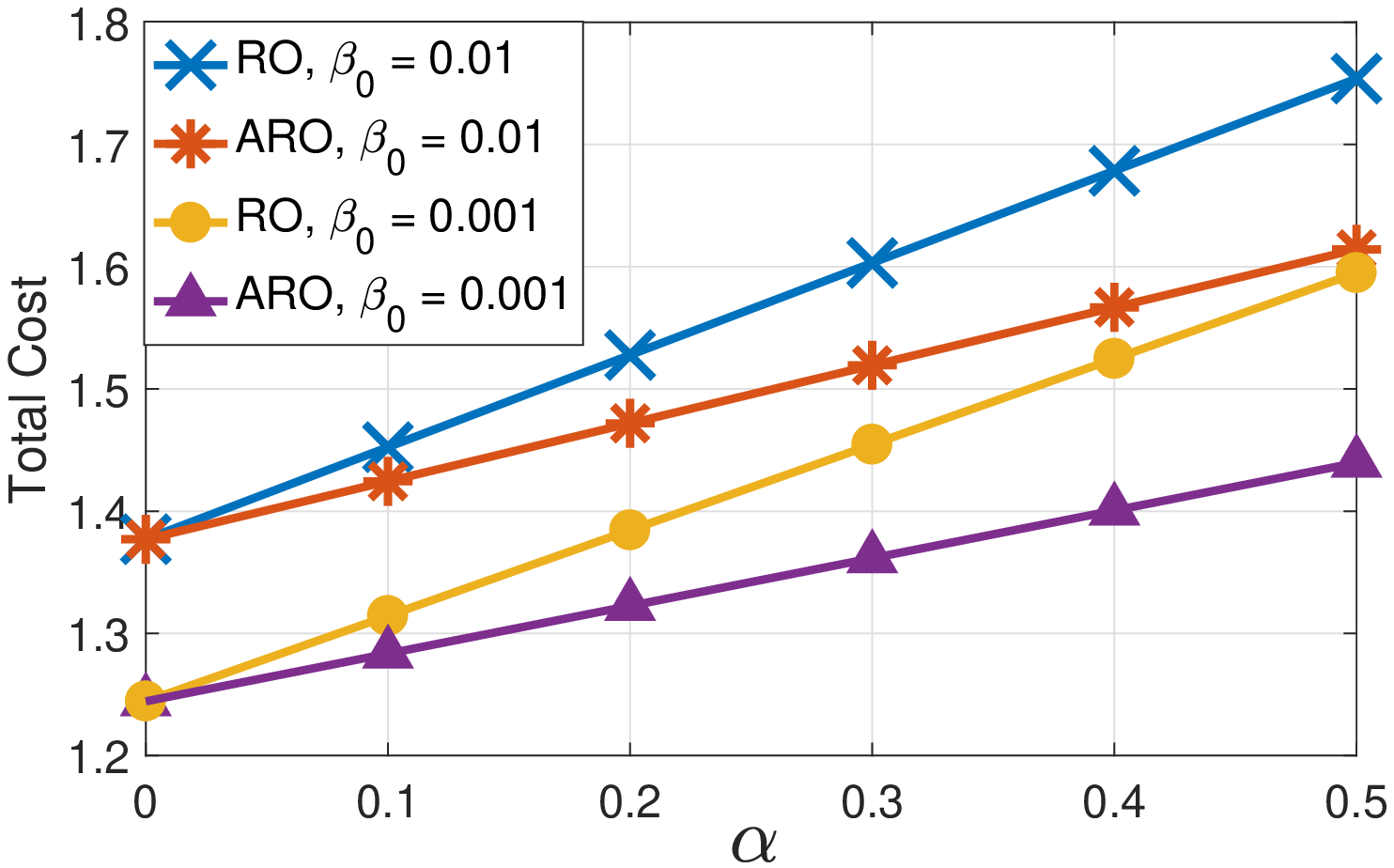}
	    \label{costroaro}
	}   \hspace*{-2.1em} 
		 \subfigure[Payment]{
	     \includegraphics[width=0.245\textwidth,height=0.12\textheight]{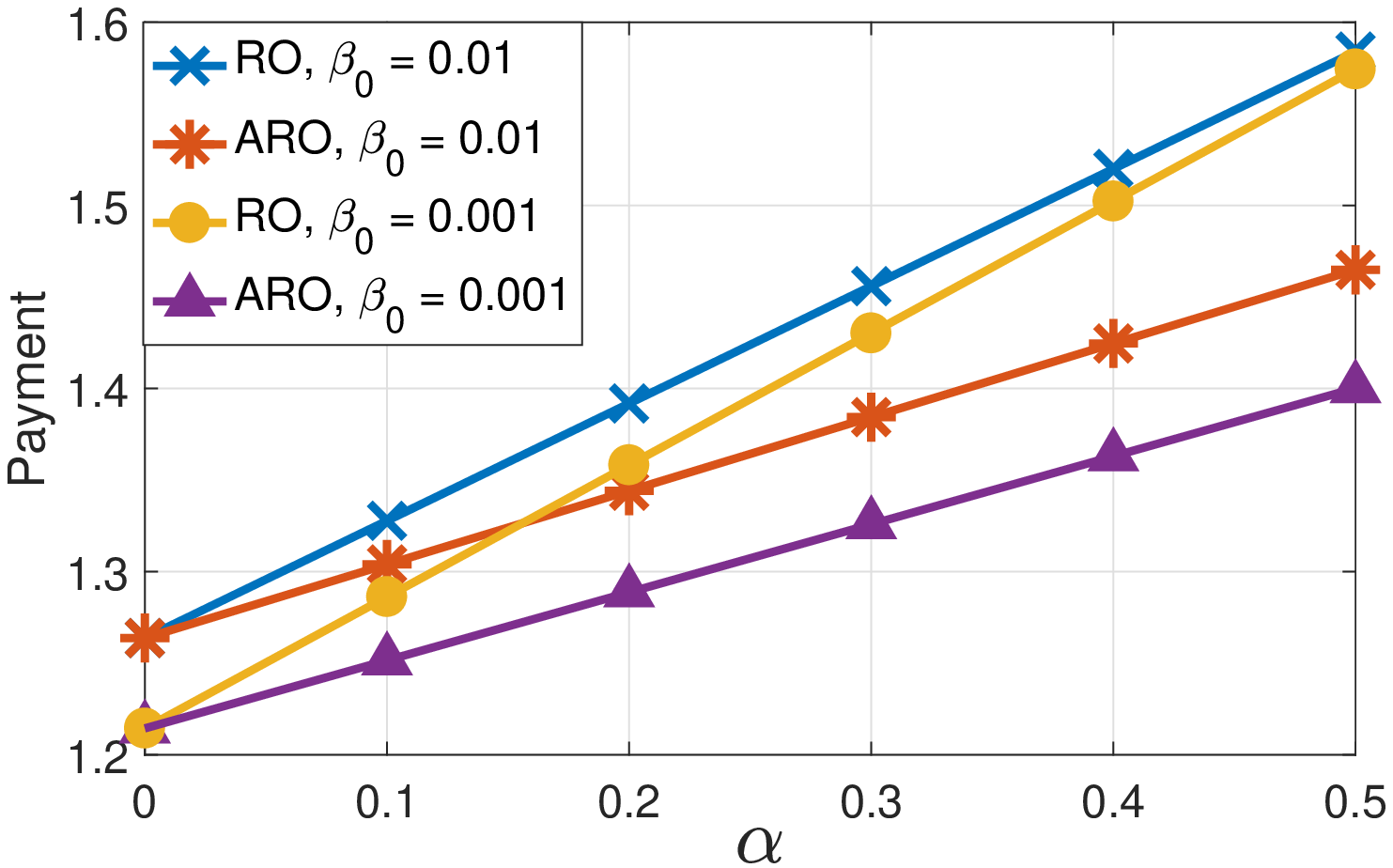}
	     \label{payroaro}
	}  \vspace{-0.2cm}
	\caption{Comparison between RO and ARO}
\end{figure}

First, we compare the performance of the traditional RO approach and the ARO approach to verify that the ARO solution is less conservative than the RO solution. Specifically, Figs.~\ref{costroaro} and \ref{payroaro} present the  costs and payments, respectively, in the RO and ARO schemes under different values of the delay penalty $\beta_0$. The payment is the total money spent for placing the service and buying resource, while the total cost is the value of the objective function that equals to the sum of the payment  and the total network delay cost. As we can observe, the total  cost as well as the payment produced by  ARO is less than those of RO. Hence, these results indicate that the ARO solution is less conservative than the RO solution.

In addition, the total cost increases as the delay penalty cost increases (i.e., the SP is more delay-sensitive and  willing to pay more to reduce the delay). Note that $\alpha$ is  the ratio between the maximum demand deviation and the forecast demand, which can be understood as the \textbf{maximum forecast error} (e.g., $\alpha$ = 0.2 implies an error of 20\% of the forecast value).   It can  be seen that the costs and payments in the robust approaches increase as the forecast error $\alpha$ increases.

When $\alpha = 0$ (i.e., no uncertainty), the ARO and RO models 
become deterministic, and the robust solutions are the same as the deterministic solution.
The  cost and payment in the deterministic model (i.e., $\alpha = 0$) are lower than those in the robust models because  the deterministic model assumes that the SP has perfect knowledge of the demand. 

\begin{figure}[h!]
		\subfigure[Demand Scale = 1]{
		  \includegraphics[width=0.245\textwidth,height=0.11\textheight]{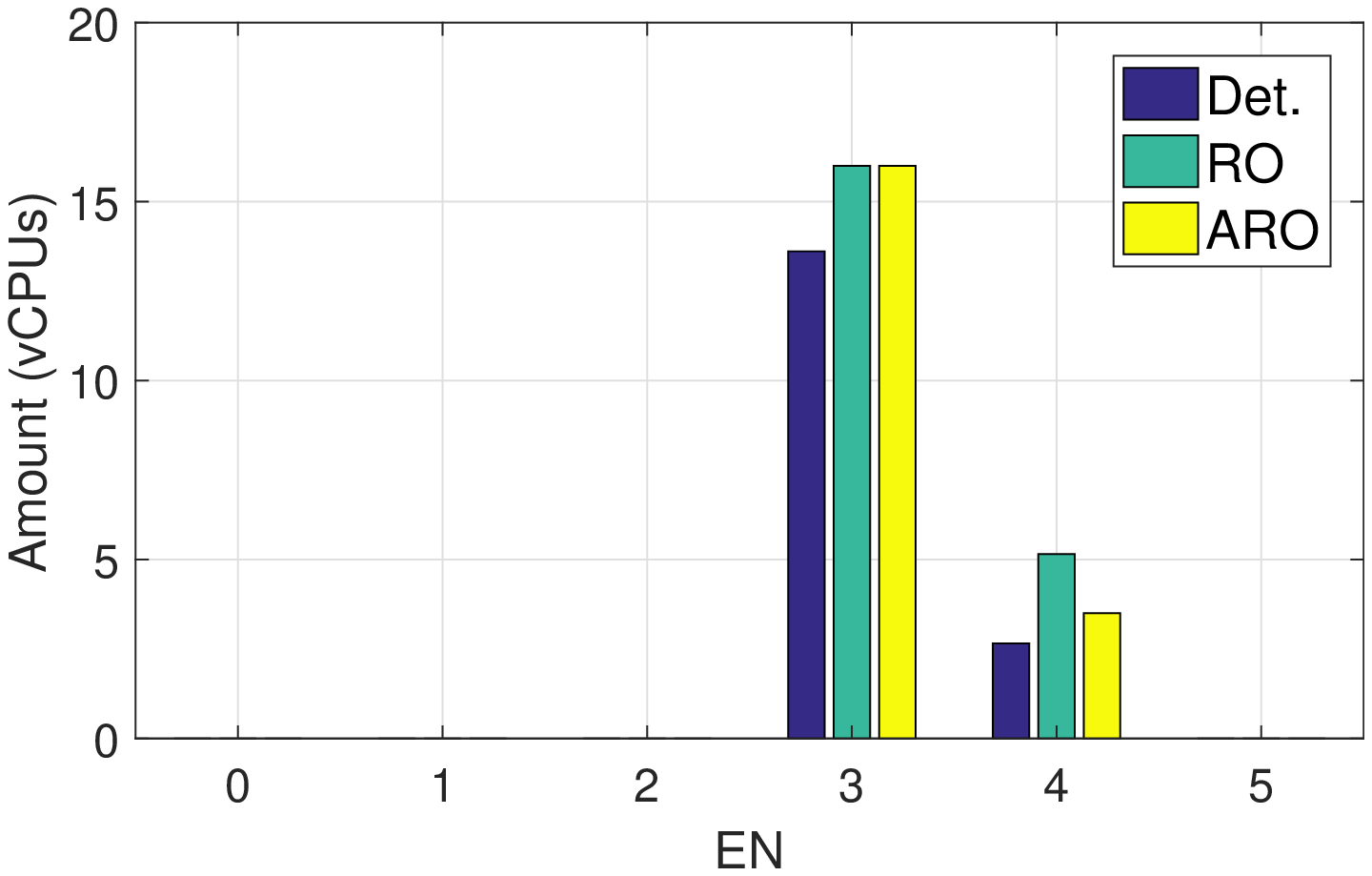}
	    \label{buyroaro1}
	}   \hspace*{-2.1em} 
		 \subfigure[Demand Scale = 2]{
	     \includegraphics[width=0.245\textwidth,height=0.11\textheight]{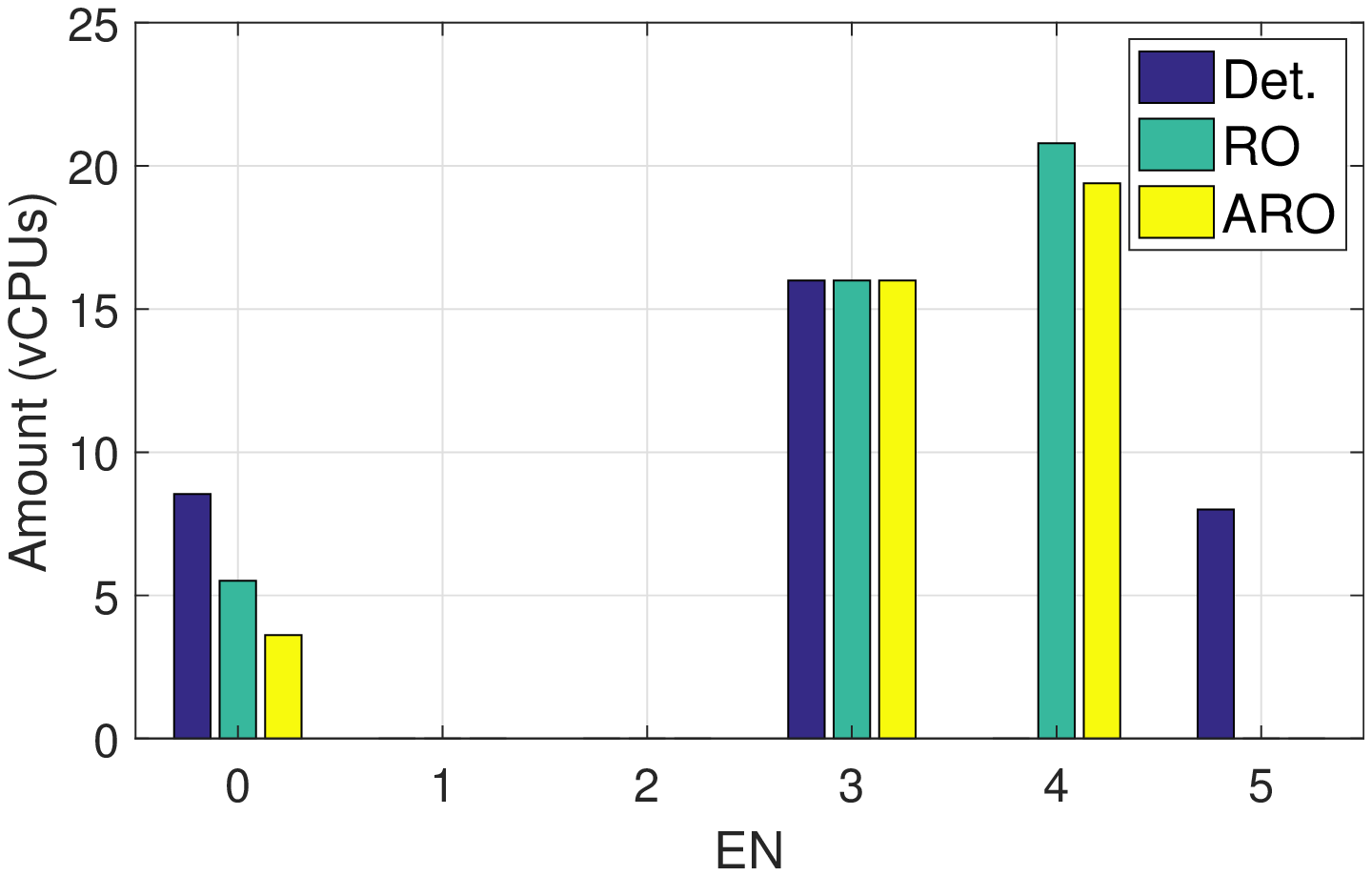}
	     \label{buyroaro2}
	}  \vspace{-0.2cm}
	\caption{Resource procurement comparison}
\end{figure}

Figs.~\ref{buyroaro1} and \ref{buyroaro2} show the amount of resources procured from the ENs and the cloud (i.e., EN 0)   in the deterministic scheme, RO scheme, and ARO scheme for a specific problem instance. The demand in each figure is equal to the demand in the base case multiplied by the demand scale factor. Clearly,  in the deterministic model, the SP buys the lowest amount of resources since the demand is assumed to be exactly known. 
It can also be observed that the amount of procured resources in the ARO model is less than that in the RO model.

Note that the   resource procurement solution depends on 
various factors such as the compute and storage prices at the cloud and the ENs, the capacity of each node, the  network topology,  the demand distribution in different areas, and the delay penalty. 
In general, the SP  prefers to buy resources from ENs with lower  prices and/or closer to high-demand areas. When the delay penalty parameter $\beta_0$ is small, the SP tends to purchase resources from low-price nodes, which may  be far from high-demand areas. When $\beta_0$ is sufficiently large, the SP tends to buy resources from ENs in or near high-demand areas to reduce the overall delay, even though  these ENs can be more expensive. It is because the delay penalty can outweigh other costs in this case. 

For the  problem instance in Figs.~\ref{buyroaro1} and \ref{buyroaro2}, EN1 and EN2 have the highest resource prices while the prices at EN3 are lowest. Furthermore, the demand is concentrated in  the areas close to EN3--EN5. Hence,  the SP prefers to buy more resources from EN3 and EN4, which offer the cheapest  prices. In the normal demand condition in Fig. \ref{buyroaro1}, the robust solutions suggest  the SP to purchase all the compute resource of EN3, whose capacity is 16 vCPUs, 
and the remaining  from EN4. In the deterministic model, the SP knows the exact demand in each area, and thus buys less resources. Also, the SP  does not buy all  resources of the cheapest node (EN3), but procures resources from EN4 to serve the demand near EN4 and EN5.

Although the cloud resource price is cheaper than the edge resource prices, the SP does not buy cloud resources due to the higher delay penalty cost at the cloud and the sufficiently low resource prices at EN3 and EN4. It is worth emphasizing that this does not always hold true and depends on the specific problem instance. For example, for the same problem instance except that the delay penalty parameter $\beta_0$ is set to be  smaller, the SP may purchase a certain amount of computing resources from the cloud.

When the demand doubles as in Fig.~\ref{buyroaro2}, in the robust schemes, the SP buys sufficient resources from EN3 and EN4 to serve the demand in the areas near EN2--EN5. In the deterministic model, the demand is assumed to be known precisely. Thus, the SP installs the service and buys resources at EN5 even though the prices at EN5 are higher than those at EN3 and EN4. The difference between the robust solutions and the deterministic solution at EN4 and EN5 is because the robust solutions need to take the demand uncertainty into account. When $\beta_0$ is not too large and the resource prices at EN1 and EN2 are highest, the total resource and delay cost by serving requests at the cloud is lower than the total cost at EN1 and EN2. Hence, it is more economical for the SP  to purchase  resources from the cloud instead of EN1 and EN2.

\subsubsection{Comparison between RO, ARO, and  deterministic approaches in the operation stage}
\label{operationCompare}
It is worth emphasizing that the main goals of the RO, ARO, and deterministic models presented throughout the paper are  to identify the placement and sizing decisions  \textit{before} knowing the actual demand. The robust approaches aim to hedge against any realization of the demand (i.e., robust against worst-case scenarios), while the deterministic approach  uses the forecast demand to determine the optimal placement and sizing solution. However, the actual demand does not necessarily coincide with either the forecast demand or the worse-case demand scenario.
Thus, it is important to evaluate the performance of these approaches in the actual operation stage when the demand is disclosed. 

 Indeed, the  cost in the deterministic model should be acquired by solving it based initially on the forecast demand and then re-evaluating it under demand uncertainty. Specifically, we first solve the deterministic model using the forecast demand to obtain the optimal  decision $(y^*, z^*)$. When the actual demand is revealed, the SP then solves the workload allocation problem 
 using $y^*$ as an input. Since $y^*$ 
may not be sufficient to serve the actual demand, we allow dropping requests in the operation stage at a penalty cost $v^{\sf p}$ per percentage of unserved requests (see \textit{Appendix} \ref{DetermiA} for more details).

Thus, the actual cost of the deterministic model is the sum of the service placement and resource procurement costs before knowing the demand (i.e.,  first-stage cost) and the actual delay cost in the operation stage (i.e.,  second-stage cost). 
We apply  similar procedures to the robust approaches since the actual demand is  rarely the same as the worst-case demand scenario. In particular, the service placement and resource procurement costs are the costs in the first-stage before the demand is revealed. Given these decisions, the delay cost in the second-stage is re-computed when the actual demand is disclosed.

\begin{figure}[h!]
		\subfigure[Average ]{
		  \includegraphics[width=0.245\textwidth,height=0.11\textheight]{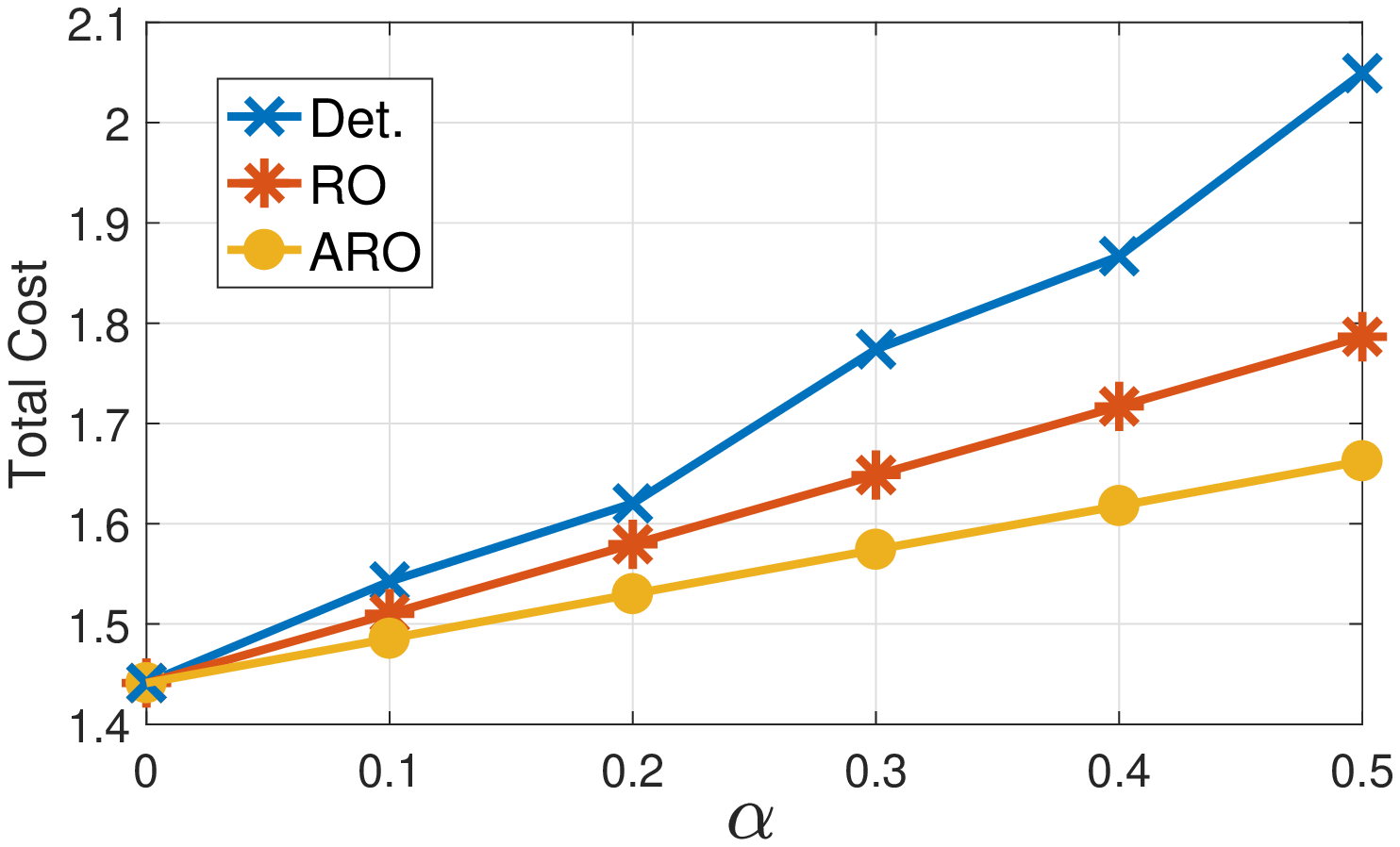}
	    \label{aveCost}
	}   \hspace*{-2.1em} 
		 \subfigure[Worst-case]{
	     \includegraphics[width=0.245\textwidth,height=0.11\textheight]{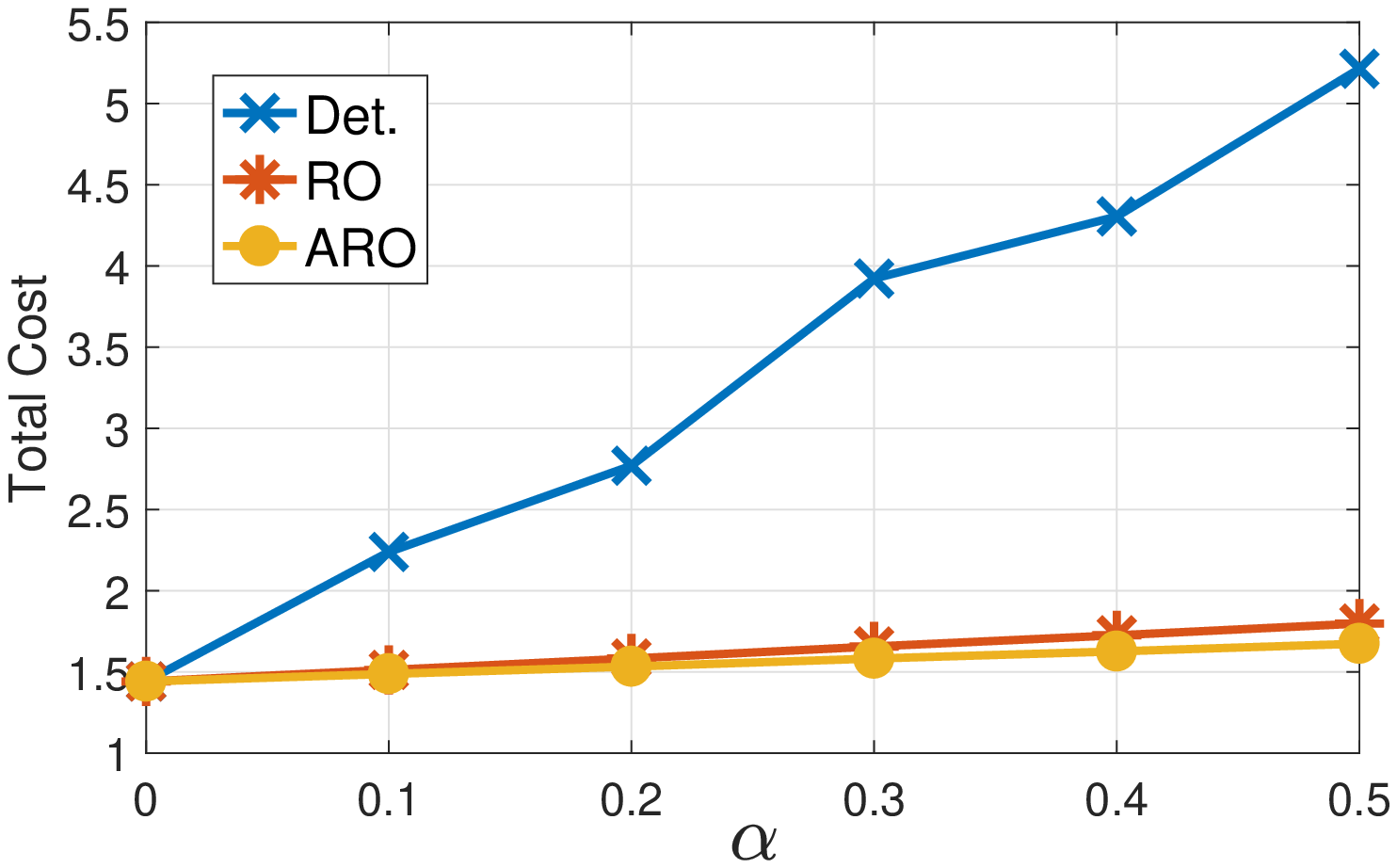}
	     \label{worstCost}
	}  \vspace{-0.2cm}
	\caption{Cost comparison under actual demand}
\end{figure}

We now compare the performance of
the deterministic and robust approaches in terms of average cost and worst-case cost. First, we generate 100 demand scenarios within the uncertainty set. 
The average cost is the sum of the second-stage cost, which is averaged over 100 scenarios, and the first-stage provisioning cost. For the worst-case cost, the second-stage cost is the cost in the worst-case demand scenario (i.e., the scenario giving the maximum second-stage cost).

\begin{figure}[ht!]
	\centering
		\includegraphics[width=0.30\textwidth,height=0.125\textheight]{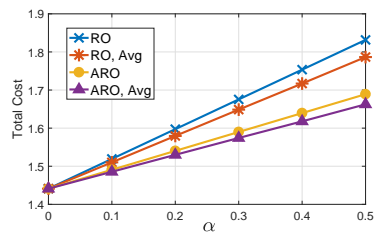}
			\caption{Robust cost and actual cost}
	\label{avgroaro}
\end{figure} 

The average cost  and the worst-case cost comparisons among these approaches are shown in Figs.~\ref{aveCost} and \ref{worstCost}, respectively.
Here, we set $v^{\sf p} = 40$ (e.g.,  the penalty will be $0.4$ if 1\% of requests are unserved).
Since the deterministic method does not consider demand uncertainty, its cost 
is significantly higher than those of the robust schemes, especially in the worst-case scenario. 
Also, because the realized demand in the operation stage is usually not the worst-case demand scenario, the actual costs of the robust solutions are  considerably lower than the optimal values of the objective functions (\ref{ROobj}) and (\ref{AROori}) in the robust models, as can be seen in Fig.~\ref{avgroaro}. This figure further confirms that ARO is less conservative than RO.

\begin{figure}[h!]
		\subfigure[Varying $\alpha$ and $\Gamma$ ]{
		  \includegraphics[width=0.245\textwidth,height=0.11\textheight]{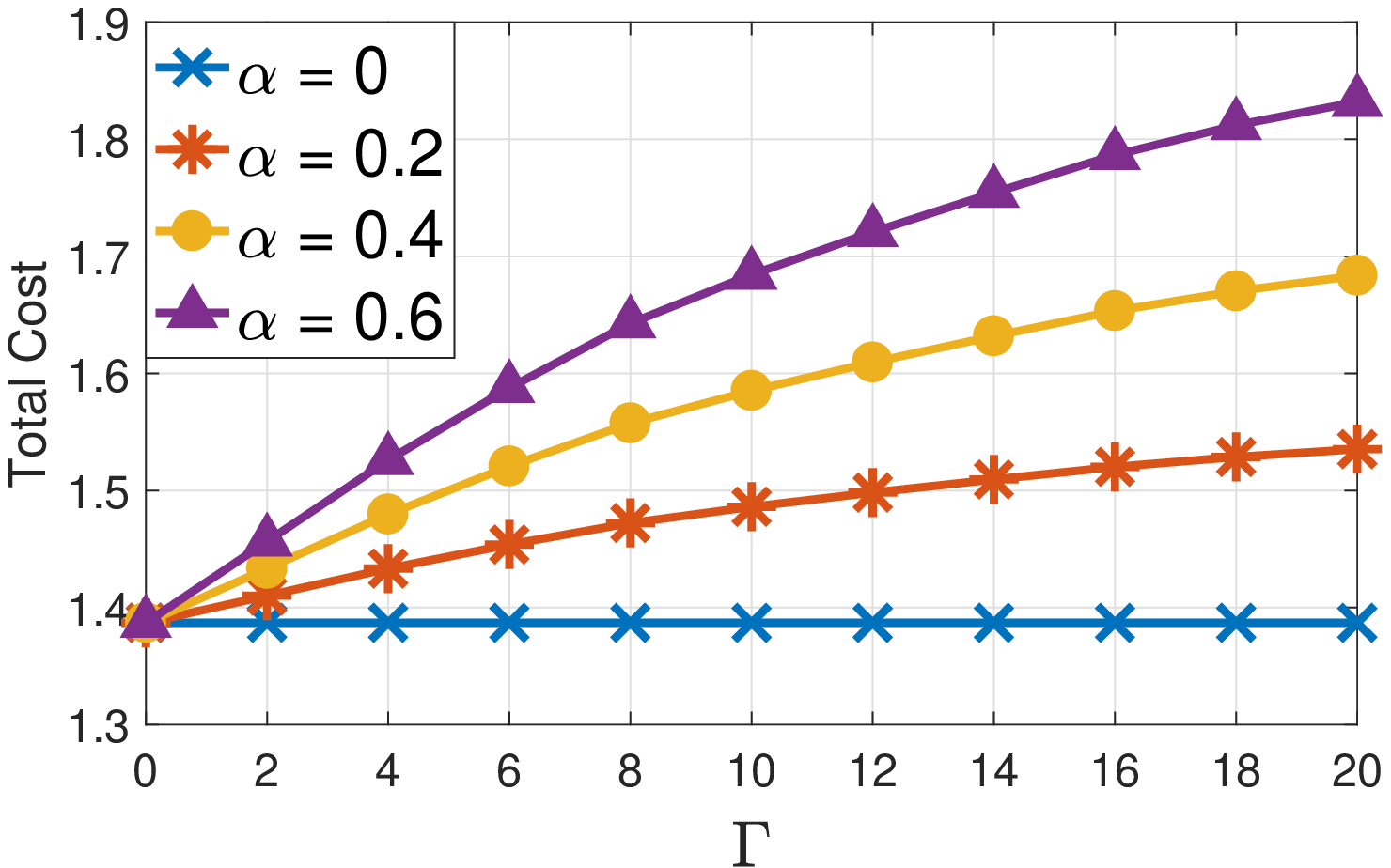}
	    \label{alphaga}
	}   \hspace*{-2.1em} 
		 \subfigure[Varying $\beta_0$ and $\Gamma$]{
	     \includegraphics[width=0.245\textwidth,height=0.11\textheight]{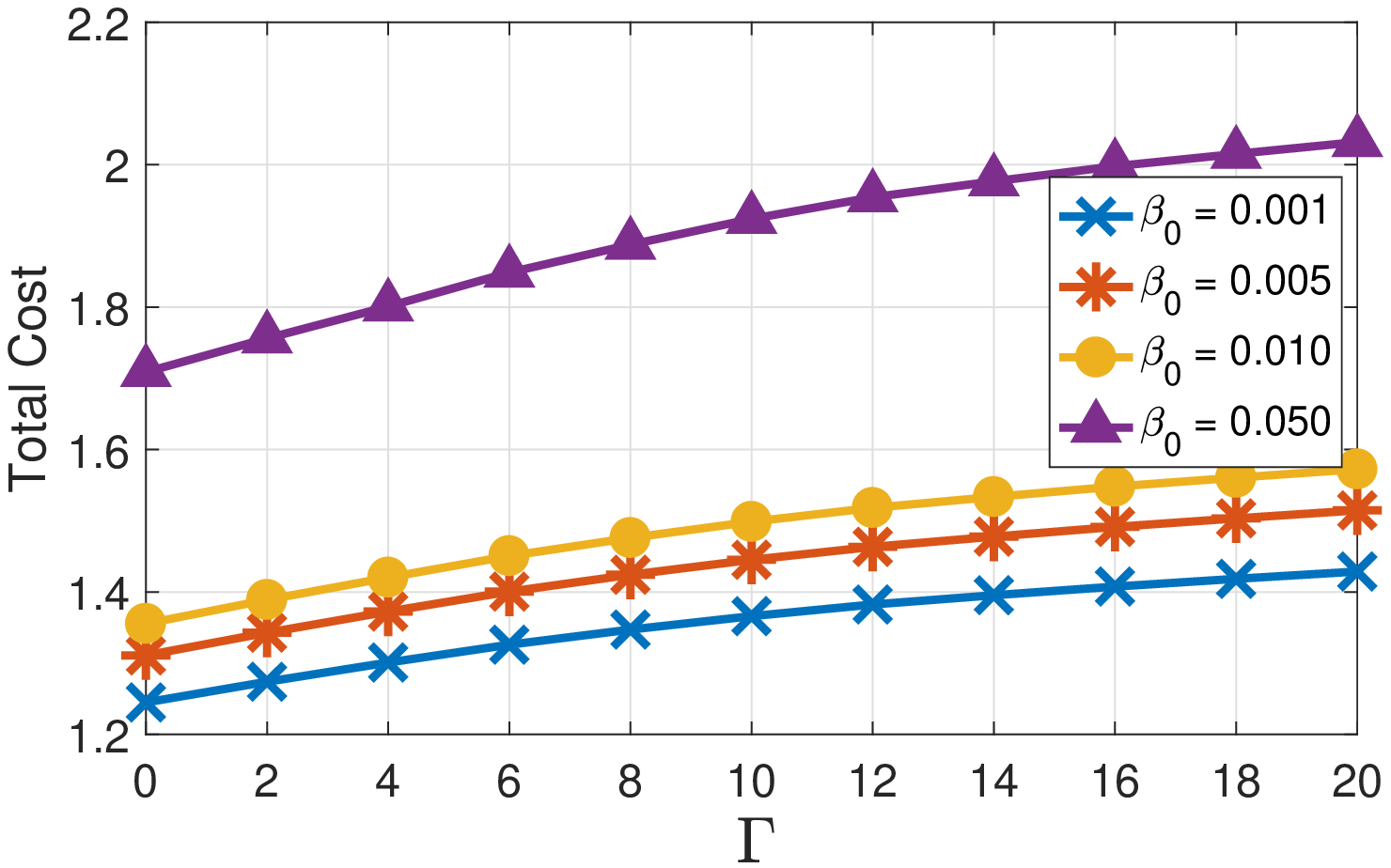}
	     \label{betaga}
	}  \vspace{-0.2cm}
	\caption{The impact of the uncertainty set on the system performance}
\end{figure}

\subsubsection{Sensitivity Analysis} We now study the impacts of different design parameters on  the system performance.
First, by definition, the uncertainty set is characterized by the maximum forecast error $\alpha$
and the uncertainty budget $\Gamma$. Figs.~\ref{alphaga} and \ref{betaga} show the impact of the uncertainty set on the optimal solution. As expected, the total cost increases as the uncertainty set $\mathcal{D}$ enlarges (i.e., $\alpha$ increases and/or $\Gamma$ increases). Note that the maximum value of $\Gamma$ is 20 since we have 20 APs in the base case. Also, 
Fig.~\ref{betaga} suggests that the SP can lower the total cost  by reducing  the delay penalty parameter $\beta_0$. 

\begin{figure}[t!]
		\subfigure[M = 20 , varying N and $\Gamma$]{
		  \includegraphics[width=0.245\textwidth,height=0.11\textheight]{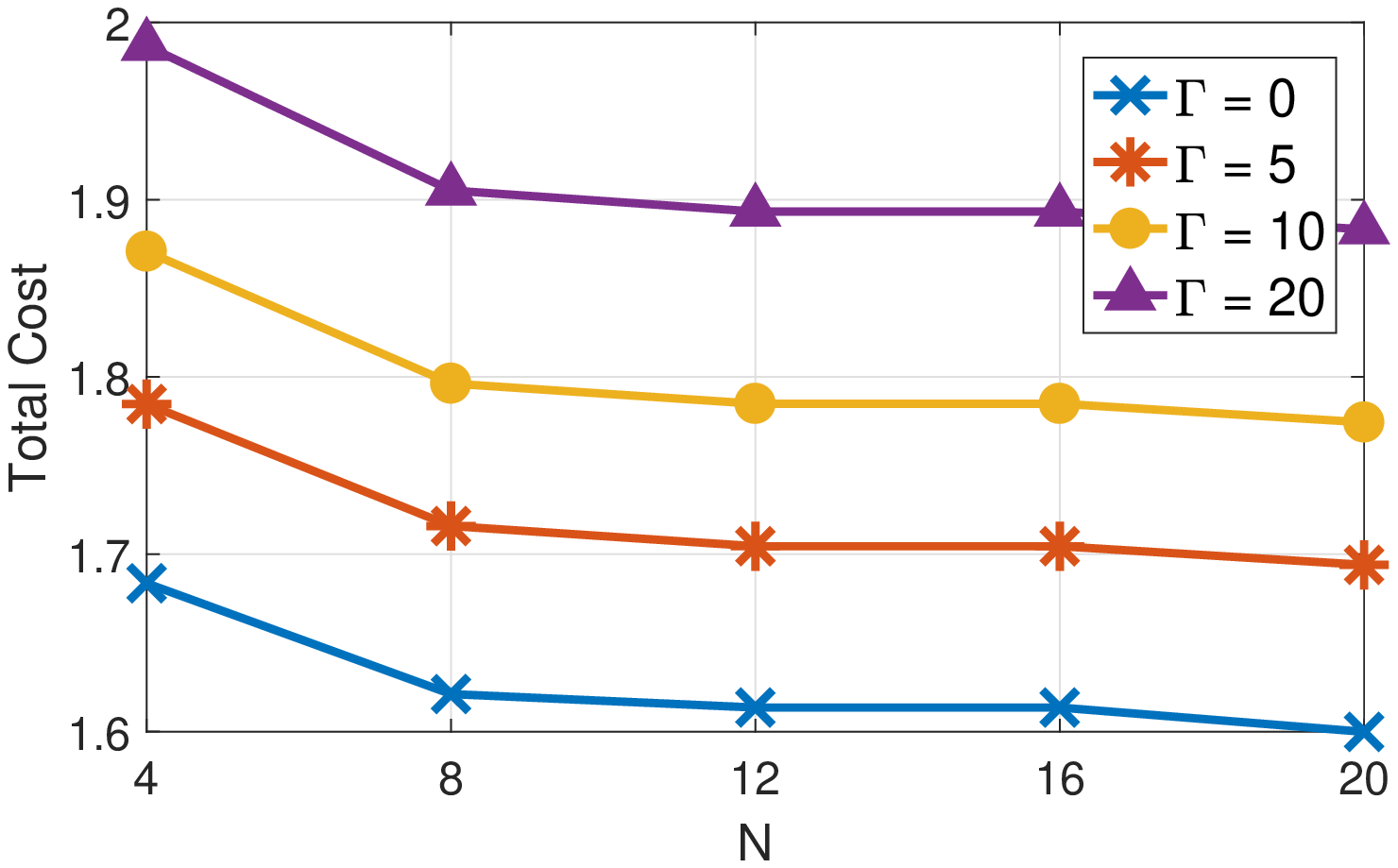}
	    \label{Mga}
	}   \hspace*{-2.1em} 
		 \subfigure[ N = 20, varying M and $\Gamma$]{
	     \includegraphics[width=0.245\textwidth,height=0.11\textheight]{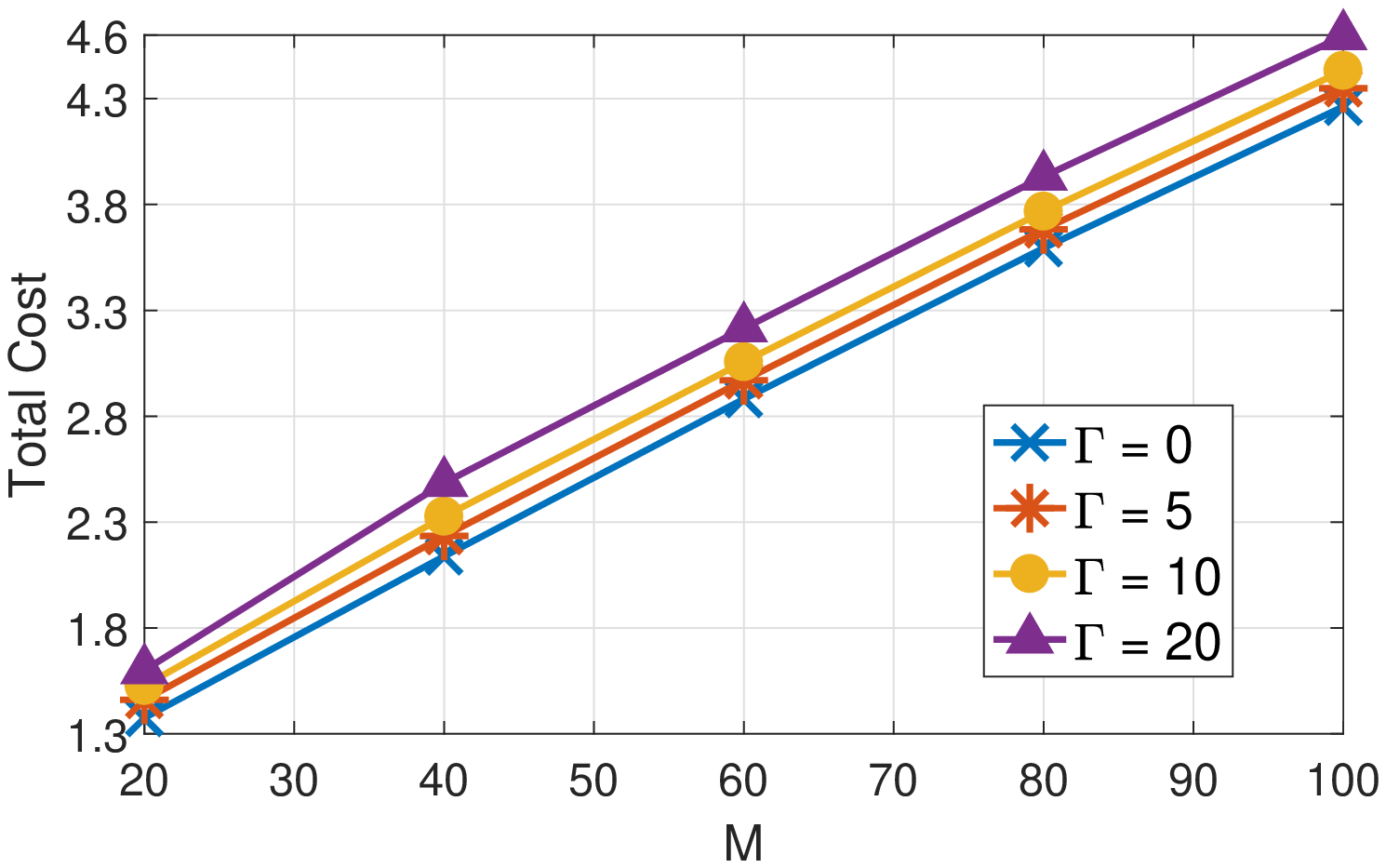}
	     \label{Nga}
	}  \vspace{-0.2cm}
	\caption{The impact of M and N on the system performance}
\end{figure}

\begin{figure}[t!]
		\subfigure[$\alpha = 0.6, \beta_0 = 0.01$]{
		  \includegraphics[width=0.245\textwidth,height=0.11\textheight]{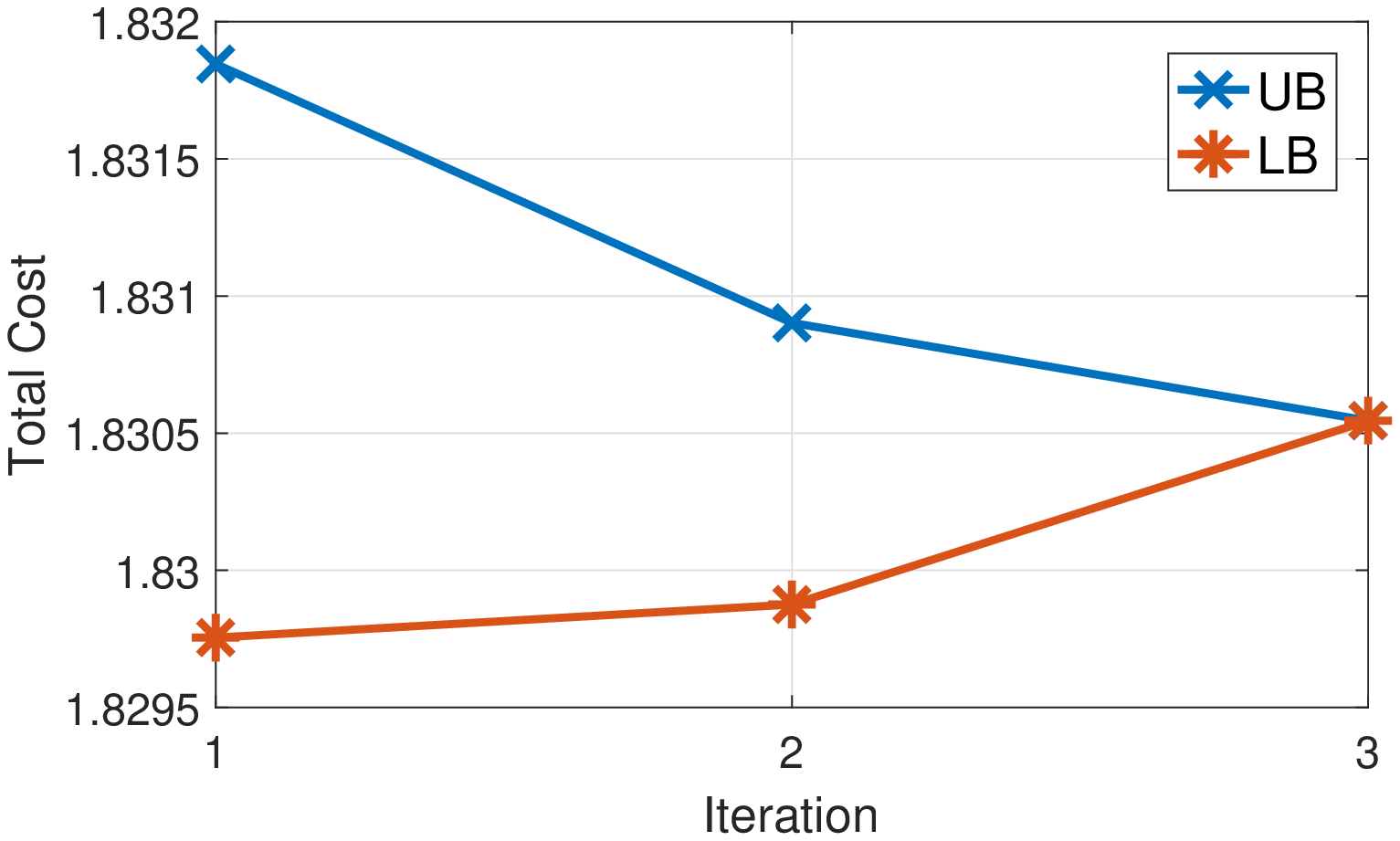}
	    \label{fig:conv1}
	}   \hspace*{-2.1em} 
		 \subfigure[ $\alpha = 0.3, \beta_0 = 0.1$]{
	     \includegraphics[width=0.245\textwidth,height=0.11\textheight]{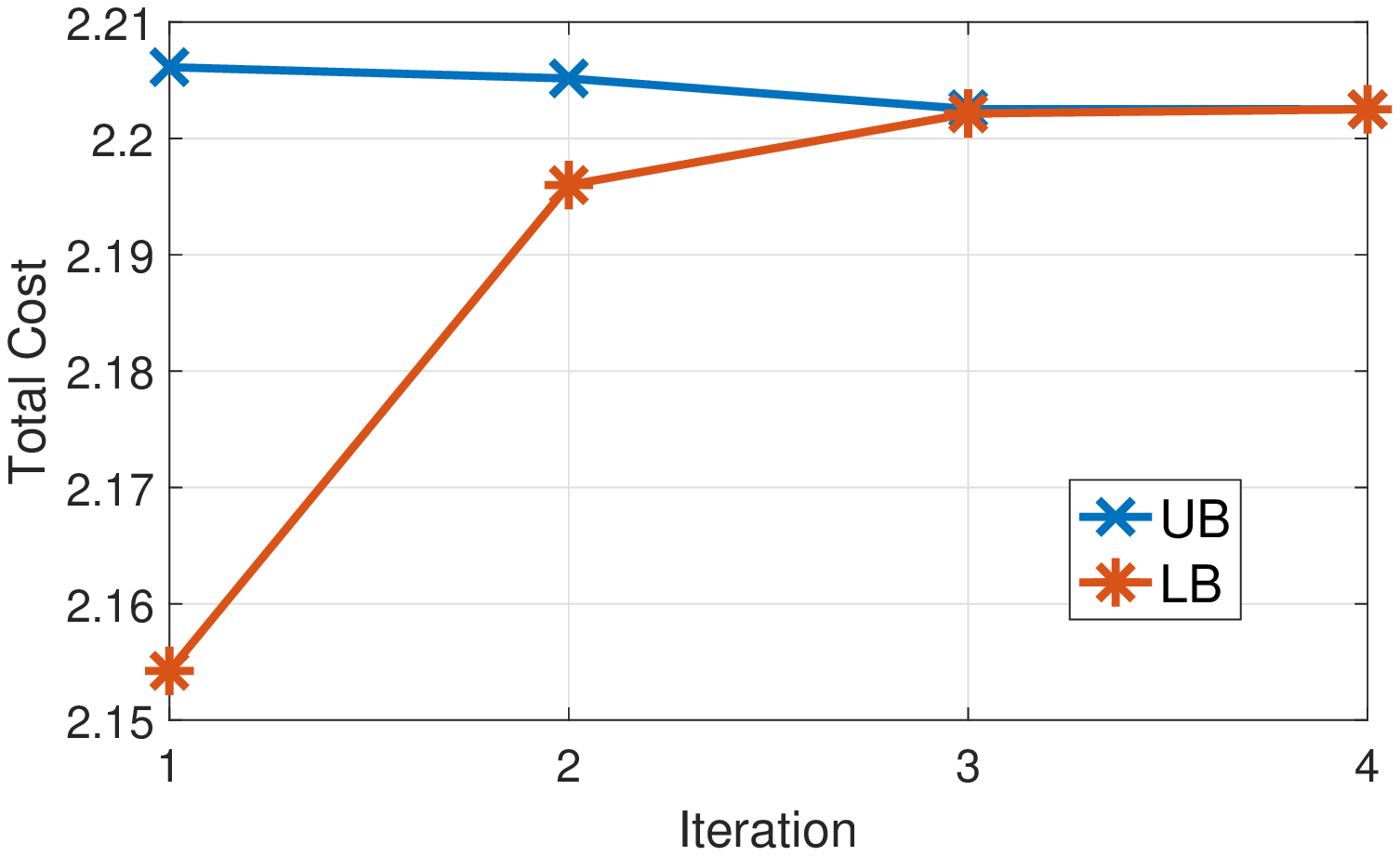}
	     \label{fig:conv2}
	}  \vspace{-0.2cm}
	\caption{Convergence property}
\end{figure}

Figs.~\ref{Mga} and \ref{Nga} illustrate the impact of number of ENs and number of APs, respectively, on the system performance. It is easy to see that when there are more ENs, the SP has more options to buy edge resources and allocate workload to closer ENs. Hence, the total cost of the SP decreases as N increases. Similarly, the total cost increases when there are more APs due to increasing total workload. Finally, the convergence property of the proposed algorithm is presented in Figs.~\ref{fig:conv1} and \ref{fig:conv2} for certain problem instances. It can be seen that the algorithm converges very quickly towards the optimal solutions. Indeed, we conducted extensive numerical experiments which show that the algorithm typically converges in a few iterations (even just one or two iterations in some cases).

\section{Related Work}
\label{rel} 

Various aspects of EC have been studied  over the last few years.
A majority of the previous work has focused on the joint allocation of communication and computational resources for task offloading in wireless networks \cite{ymao17}. 
In \cite{hzha17},
Stackelberg game and matching theory are combined  to tackle the fog resource allocation problem.
Reference \cite{duong3} presents a primal-dual method for online matching edge resources to different service providers  to maximize  system efficiency. 
A cloudlet load balancing problem is formulated in \cite{mjia18} to minimize the maximum response time of  offloaded tasks. 
R.  Deng \textit{et al.} \cite{rden16} propose a novel workload allocation model in a hybrid cloud-fog system to minimize   energy  cost under latency constraints. 
In \cite{duong1,duong2}, a market equilibrium approach is employed  to fairly and efficiently allocate  edge resources
to  competing 
services. 
Unlike  these  works,  we  focus  on  the  joint  edge  service placement  and  workload  allocation  problem.

 A growing literature has focused on 
 the optimal placement and activation of ENs. This  line  of  research  is  complementary to  our  work  where  we  examine  the  optimal  placement of  edge  services  and  assume  the  set  of  ENs  is  given.
In \cite{zxu16}, 
the authors
jointly optimize
cloudlet placement and workload allocation to minimize the average network delay between mobile users and the cloudlets by placing a given number of cloudlets to some strategic locations.
This model is extended in \cite{mjia17} 
to capture both network delay and computing delay at the cloudlets using queuing models.
In \cite{lma17}, L. Ma \textit{et al.}  formulate a cloudlet placement and resource
allocation problem, with the goal of minimizing the number
of cloudlets while respecting the access delay requirements of  users.

A ranking-based near optimal algorithm
is presented in \cite{lzha18} for efficiently deploying cloudlets among numerous APs in an IoT network.
 Reference \cite{syan19} aims 
 to minimize the overall energy consumption subject to delay constraints by intelligently placing cloudlets on the network and allocating  tasks
to cloudlets and the public cloud. 
In \cite{aces17}, A. Ceselli \textit{et al.} introduce a mobile edge cloud network planning framework that simultaneously considers cloudlet placement, assignment of APs to cloudlets, and traffic routing  from and to the cloudlets, with the goal of minimizing the total installation costs of all network nodes. 
A  MINLP formulation 
 is proposed in  \cite{smon18}  to
determine optimal locations for cloudlet placement  with minimum installation cost considering the capacity and latency constraints.
Reference \cite{fhai19} presents a multi-objective optimization framework to minimize the service delay and the cloud's load by simultaneously identifying the optimal location, capacity, and number of ENs, as well as the optimal links between the ENs and the cloud.

Recently, service  placement in EC has attracted a lot of attention in the literature. 
In \cite{lyan16}, a unified service placement and request dispatching framework is proposed to optimize the tradeoffs between the average latency of users' requests and the cost of service providers. In \cite{dzen16}, the authors present an application image placement and task scheduling problem in a fog network with dedicated storage and computing servers to minimize the makespan. A constant-factor  approximation algorithm is introduced in \cite{spas19}  to find a feasible service placement that maximizes the total user utility considering the heterogeneity of ENs 
and user locations. 
  In  \cite{ryu19}, R. Yu \textit{et. al.} consider an IoT application provisioning problem that jointly optimizes application placement and data routing to support all data streams with both bandwidth and delay
guarantees.

In \cite{qfan18}, a joint  application placement and workload allocation  scheme is presented to minimize the response time of IoT application requests.

The work  \cite{ayou19} introduces a fog service provisioning framework that dynamically deploys and releases applications on fog nodes to satisfy low latency and QoS requirements of the applications while minimizing the total system cost. The service entity placement problem for social virtual reality applications is examined in \cite{lwan18} to minimize the total system cost, including the cloudlet activation cost, the placement cost, the proximity
cost, and the colocation cost, for deploying these applications at the edge. Joint optimization  of access point selection and service placement is addressed in \cite{bgao19} to enhance user QoS by balancing the access delay, communication delay, and switching delay.

In \cite{kpout19}, the authors jointly optimize service placement and  request  routing  in  MEC-enabled  multi-cell networks 
to reduce the load of the centralized cloud considering the limited storage capacities of ENs and asymmetric bandwidth requirements of the services.
In \cite{vfar19}, a two-time-scale optimization framework is proposed
to optimize service placement and request scheduling under multi-dimensional resource and budget constraints, in which request scheduling occurs at a smaller time scale and service placement  occurs at a larger scale to reduce system instability. 
Most of the existing work on edge service  placement  does  not  consider  uncertainties.

The resource allocation and provisioning problem under uncertainty in cloud/edge computing has also been studied in recent literature.
In \cite{touy19},  T.  Ouyang \textit{el. al.} formulate the dynamic service placement problem as a contextual multi-armed bandit problem
and propose  a Thompson-sampling based online learning algorithm to assist users to select an EN for offloading  considering the tradeoff between latency and service migration cost.
In the same line of research, due to the unknown benefit of placing edge service in a specific site, 
a combinatorial contextual bandit learning problem is presented in  \cite{lche18} to help an application provider decide the optimal set of ENs to host its service under the budget constraint.

In \cite{scha12, smir20}, the authors employ
the scenario-based stochastic programming approach to tackle different cloud resource provisioning problems 
that aims to minimize the total 
resource provisioning cost. Similarly, reference \cite{hbad20} formulates an
energy-aware edge service placement as a multi-stage stochastic program with the objective of maximizing the QoS of the system under the limited energy budget of edge servers.
In \cite{scha10,rkae13}, the authors propose different robust cloud resource provisioning formulations using the standard RO method.
Also, RO is  utilized in \cite{yli18} to jointly optimize radio  and  virtual machine resources in mobile edge computing.

Different from the existing literature, we formulate a new edge service placement problem  from the perspective of a SP. Also, to deal with demand uncertainty, we propose a novel two-stage 
 RO model for joint optimization of edge service placement, sizing,
 and workload allocation.
To the best of our knowledge, this is the first work
that employs two-stage adaptive robust optimization to tackle the edge service placement and resource procurement problem.

\section{Conclusion and Future Work}
\label{conc}
In this paper, we introduced a novel two-stage RO model to help a SP decide an optimal  service placement and sizing solution that can hedge  against all possible realizations of the uncertain traffic demand within an   uncertainty set. 
Given the placement and sizing decision in the first-stage, 
the workload  allocation decisions are made in the second-stage after the uncertainty is revealed.
The proposed robust model enables the SP to balance the tradeoff between the total operating cost and the service quality while taking demand uncertainty into account.
Extensive numerical results were presented to
demonstrate the advantages of the proposed scheme,
which is less conservative compared to the RO approach and more robust compared to the deterministic approach.

There are several interesting directions for future work. First, we would like to study more efficient methods to speed up the computation of both the master problem and the subproblem. Second, instead of polyhedral sets, we would like to explore techniques to tackle more sophisticated and/or data-driven uncertainty sets (e.g., one that can better capture the highly dynamical and correlated uncertainty parameters). We also  plan to extend the proposed model to the network slicing problem where a network operator 
optimizes the planning and operation of their edge network to serve multiple services with uncertain demand and different characteristics.
 Finally, we are  interested in studying the impact of the placement and sizing decisions of the SP on the resource prices when there are multiple SPs in the system.

\bibliographystyle{IEEEtran}


\appendix
\subsection{Deterministic Formulation}
\label{DetermiA}
The deterministic formulation of the service placement and sizing problem is the following MILP problem:
\beqn
\label{eq:deterObj}
 \underset{x,y,z}{\text{min}}~  \sum_j f_j (1 - z_j^0) z_j + \sum_j s_j z_j +  p_0  y_0  +  \sum_j p_j y_j    \nonumber \\ 
 + \beta  \Bigg\{ \sum_i d_{i,0} x_{i,0} + \sum_{i,j} d_{i,j} x_{i,j} \Bigg\} 
\eeqn
subject to 
\beqn
\sum_j z_j \geq r^{\sf min} \\ 
y_j \leq z_j C_j, ~~ \forall j \\
\sum_j f_j (1 - z_j^0) z_j + \sum_j s_j z_j + p_0  y_0  + \sum_j p_j y_j  \leq B \\
w \sum_i x_{i,0} \leq y_0\\ 
w \sum_i x_{i,j} \leq y_j, ~~ \forall j \\
x_{i,0}  + \sum_j x_{i,j} = \lambda_i, ~~ \forall i \\
d^{\sf avg} = \frac{\sum_i d_{i,0} x_{i,0} + \sum_i  \sum_j d_{i,j} x_{i,j}  }{\sum_i \lambda_i^{\sf f}} \\ 
d^{\sf avg}    \leq D^{\sf m} \\ 
z_j \in \{0,~1\};~~y_0 \geq 0;~~y_j \geq 0,~\forall j \\ 
\label{eq:deterLast} 
x_{i,0} \geq 0,~\forall i;~~x_{i,j} \geq 0,~ \forall i,j.
\eeqn

In the deterministic algorithm in the simulation section, the SP first solves this deterministic problem, using the forecast demand $\lambda^{\sf f}$, to obtain the optimal values of $z$ and $y$ (i.e., the amount of resource that the SP buy from the cloud and the ENs).  Note that the SP can decide the optimal workload allocation (i.e., $x$) after observing the actual workload. However, the SP does not know the actual demand when making the resource procurement decision. Thus, the procured resources may not be able to serve the actual demand when the actual demand is much higher than the forecast value, which makes the workload allocation problem become infeasible.

Therefore, we define $x_i^{\sf D}$ as the number of requests at AP $i$ that are unserved (i.e., dropped requests) in the operation stage. Let $ x^{\sf D} = \{x_1^{\sf D}, x_2^{\sf D},\ldots, x_M^{\sf D} \}$. Also, denote by $v^{\sf p}$ the penalty for unserved requests. Let $Drop$ be the ratio between  the total number of unserved requests and the total number of requests, i.e., $Drop = \frac{\sum_i x_i^{\sf D}}{\sum_i \lambda_i}$, where $\lambda_i$ is the actual workload arriving at AP $i$ in the operation stage.
Given the realization of  demand $\lambda = \{ \lambda_1, \lambda_2, \ldots, \lambda_M \}$,  the SP solves the following problem to determine the optimal workload allocation: 
\beqn
\label{obj:cor}
 \underset{x,x^{\sf D}}{\text{min}}~   \beta  \Bigg\{ \sum_i d_{i,0} x_{i,0} + \sum_{i,j}   d_{i,j} x_{i,j} \Bigg\} + v^{\sf p} \frac{\sum_i x_i^{\sf D}}{\sum_i \lambda_i}
\eeqn
subject to 
\beqn
w \sum_i x_{i,0} \leq y_0\\ 
w \sum_i x_{i,j} \leq y_j, ~~ \forall j, \\ 
x_{i,0}  + x_i^{\sf D} + \sum_j x_{i,j}  = \lambda_i, ~~ \forall i, \\ 
\label{eq:newDelay}
 \sum_i d_{i,0} x_{i,0} + \sum_{i,j} d_{i,j} x_{i,j}    \leq D^{\sf m} \sum_i \Big( x_{i,0} + \sum_j x_{i,j} \Big) \\ 
z_j \in \{0,~1\};~~y_0 \geq 0;~~y_j \geq 0,~\forall j \\ 
\label{eq:corlast}  
x_{i,0},~ x_i^{\sf D} \geq 0,~\forall i;~~x_{i,j} \geq 0,~ \forall i,j,
\eeqn
where $y$ is the optimal solution to the problem  (\ref{eq:deterObj})-(\ref{eq:deterLast}). Note that the maximum average delay constraint (\ref{eq:newDelay}) is enforced for served requests only.

For the performance evaluation section in the paper, we compare the performance of the RO, ARO  and the deterministic approaches. For example, similar to the procedures above, in the ARO approach, the SP first obtains the optimal value of  $y^{\sf ARO}$ by solving the two-stage RO problem (\ref{AROori}) and use it as input to the problem (\ref{obj:cor})-(\ref{eq:corlast}) to determine the optimal workload allocation decision with each realization of the actual workload.
Note that for sufficiently large values  of the penalty parameter $v^{\sf p}$, the optimal $x_i^{\sf D}$ in this ARO approach is equal to zero for every $i$ (i.e.,  no dropped requests) since  $y^{\sf ARO}$ is robust against any demand realization in the uncertainty set.

\subsection{Robust Optimization Formulation}
\label{ROmodelA}
In the static robust service placement and sizing problem, the placement, sizing, and workload allocation decisions are made simultaneously before knowing the actual demand. Based on the RO theory \cite{RObook}, the robust service placement and sizing problem can be expressed as follows:
\beqn
\label{eq:ROobj}
 \underset{(x,y,z) \in  \mathcal{S}_{(X,Y,Z)}  }{\text{min}} ~\underset{\lambda \in \mathcal{D}}{\text{max}} ~ \sum_j f_j (1 - z_j^0) z_j + \sum_j s_j z_j +  p_0  y_0    \nonumber \\ 
   +  \sum_j p_j y_j + \beta  \Bigg\{ \sum_i d_{i,0} x_{i,0} + \sum_{i,j} d_{i,j} x_{i,j} \Bigg\} 
\eeqn
subject to 
\beqn 
\label{eq:demandRO}
x_{i,0}  + \sum_j x_{i,j} \geq \lambda_i, \quad \forall i, \lambda \in \mathcal{D}    \\ 
\label{eq:delayRO}
\sum_i d_{i,0} x_{i,0} + \sum_i  \sum_j d_{i,j} x_{i,j}   \leq D^{\sf m} \sum_i \lambda_i, \lambda \in \mathcal{D}, 
\eeqn
where
\beqn
\mathcal{D}(\lambda^{\sf f}, \hat{\lambda}, \Gamma) = \Big\{   
&&\lambda_i = \lambda_i^{\sf f} + g_i \hat{\lambda}_i,~\forall i; ~g_i \in [-1,~1],\forall i \nonumber\\ 
&& \sum_i~ | g_i | \leq \Gamma  \quad
\Big\},
\label{uncertainty_budget1}
\eeqn

\beqn
\mathcal{S}_{(X,Y,Z)} = \Bigg\{ 
&&\sum_j z_j \geq r^{\sf min}, \\ \nonumber
&&\sum_j h_j z_j  + p_0  y_0  + \sum_j p_j y_j  \leq B, \\ \nonumber
 &&y_j \leq z_j C_j, \quad \forall j, \\ \nonumber
&&w \sum_i x_{i,j} \leq y_j, \quad \forall j, \\ \nonumber
&&w \sum_i x_{i,0} \leq y_0\\ \nonumber
&&z_j \in \{0,~1\};~~y_0 \geq 0;~~y_j \geq 0,~\forall j \\ \nonumber
&&x_{i,0} \geq 0,~\forall i;~~x_{i,j} \geq 0,~ \forall i,j
\Bigg\}
\eeqn
Note that in the deterministic formulation, both equality and inequality constraints will produce the same result since equality happens at the optimality for cost minimization objective. Furthermore, equality constraint related to uncertainty is meaningless in the RO  approach \cite{RObook}. Thus, we write the workload balance constraints in form of inequality constraints as in (\ref{eq:demandRO}). 
Since the objective function (\ref{eq:ROobj}) does not contain uncertainty parameters, we can alternatively write the robust problem as:
\beqn
 \underset{(x,y,z) \in  \mathcal{S}_{(X,Y,Z)} }{\text{min}} ~ \sum_j f_j (1 - z_j^0) z_j + \sum_j s_j z_j + p_0  y_0   \nonumber \\ 
 + \sum_j p_j y_j  + \beta  \Bigg\{ \sum_i d_{i,0} x_{i,0} + \sum_{i,j} d_{i,j} x_{i,j} \Bigg\}  
\eeqn
subject to  
\beqn
&&x_{i,0}  + \sum_j x_{i,j} \geq \underset{\lambda \in \mathcal{D}}{\text{max}}~  \lambda_i \\
\label{eq:delayRO1}
&&\sum_i d_{i,0} x_{i,0} + \sum_{i,j} d_{i,j} x_{i,j}   \leq D^{\sf m} \underset{\lambda \in \mathcal{D}}{\text{min}}   \sum_i \lambda_i. \\ \nonumber
\eeqn

Consider  constraint (\ref{eq:delayRO1}). First, observe that $\underset{\lambda \in \mathcal{D}}{\text{min}} ~ \sum_i \lambda_i$ can be expressed as:
\beqn
\label{eq:objlam}
\underset{g}{\text{min}} ~ \sum_i \Big( \lambda_i^{\sf f}  + g_i \hat{\lambda}_i \Big)
\eeqn
subject to 
\beqn
\label{eq:conslam}
\sum_i~ |g_i| \leq \Gamma ;~~
|g_i| \leq 1,~~ \forall i.
\eeqn
The problem (\ref{eq:objlam})-(\ref{eq:conslam}) above is equivalent to:
\beqn
\underset{g,~t}{\text{min}}  ~\sum_i  g_i \hat{\lambda}_i 
\eeqn
subject to 
\beqn
\sum_i t_i \leq \Gamma, \quad (u) \\
-t_i + g_i \leq 0, \quad \forall i \quad (v_i) \\ 
-t_i - g_i \leq 0, \quad \forall i \quad (\gamma_i) \\ 
g_i \leq 1,\quad \forall i \quad (\mu_i) \\ 
-g_i \leq 1,\quad \forall i \quad (\sigma_i) 
\eeqn
where  $u,v,\gamma,\mu,\sigma$ are the dual variables associated with the constraints. This is a linear program. 
Based on duality in linear programming, the dual linear program of this problem is:
\beqn
\underset{u,v,\gamma,\mu,\sigma}{\text{max}}  
- u \Gamma - \sum_i  \mu_i - \sum_i \sigma_i 
\eeqn
subject to 
\beqn
v_i - \gamma_i + \mu_i - \sigma_i \leq -\hat{\lambda}_i,\quad \forall i   \\ 
u - v_i - \gamma_i \geq 0,\quad \forall i  \\
u \geq 0;~~ v_i \geq 0,~~\gamma_i \geq 0,~~\mu_i \geq 0,~~\sigma_i \geq 0, \forall i.
\eeqn
Additionally, we have:
\beqn
\label{eq:demandR}
x_{i,0}  + \sum_j x_{i,j} \geq \underset{\lambda \in \mathcal{D}}{\text{max}}~  \lambda_i = \underset{\lambda \in \mathcal{D}}{\text{max}}~ \Big( \lambda_i^{\sf f}  + g_i \hat{\lambda}_i \Big).
\eeqn
In the set $\mathcal{D}$, we have $\sum_i 1. |g_i| \leq \Gamma$ and $-1 \leq g_i \leq 1$. Hence, $\underset{\lambda \in \mathcal{D}}{\text{max}} ~g_i = \min\Big\{1, \frac{\Gamma}{1}\Big\}$. Consequently, (\ref{eq:demandR}) is equivalent to:
\beqn
\label{eq:demandR1}
x_{i,0}  + \sum_j x_{i,j} \geq   \lambda_i^{\sf f}  + \min\Big\{1, \Gamma\Big\} \hat{\lambda}_i. 
\eeqn

Finally, the static (single-stage) robust service placement and workload allocation problem can be written as follows:

\beqn
\label{eq:ROobjfinal}
 \underset{(x,y,z) \in  \mathcal{S}_{(X,Y,Z)} }{\text{min}} ~ \sum_j f_j (1 - z_j^0) z_j + \sum_j s_j z_j + p_0  y_0    \\  \nonumber
 + \sum_j p_j y_j  + \beta  \Bigg\{ \sum_i d_{i,0} x_{i,0} + \sum_{i,j} d_{i,j} x_{i,j} \Bigg\} 
\eeqn
subject to 
\beqn
 &&x_{i,0}  + \sum_j x_{i,j} \geq
\lambda_i^{\sf f}  + \min\Big\{1, \Gamma\Big\} \hat{\lambda}_i \\
\label{eq:aveDR}
&&\sum_i d_{i,0} x_{i,0} + \sum_{i,j} d_{i,j} x_{i,j}      \\ \nonumber
&& \quad \leq D^{\sf m}  \sum_i \lambda_i^{\sf f} -  D^{\sf m} \Big( u \Gamma + \sum_i  \mu_i + \sum_i \sigma_i  \Big)  \\ 
&& u \Gamma + \sum_i  \mu_i + \sum_i \sigma_i  = - \sum_i  g_i \hat{\lambda}_i  \\ 
&&\sum_i t_i \leq \Gamma  \\  
&&-t_i + g_i \leq 0, \quad \forall i  \\
&&-t_i - g_i \leq 0, \quad \forall i  \\ 
&&g_i \leq 1,\quad \forall i  \\ 
&&-g_i \leq 1,\quad \forall i \\
&&v_i - \gamma_i + \mu_i - \sigma_i \leq -\hat{\lambda}_i,\quad \forall i   \\
&&u - v_i - \gamma_i \geq 0,\quad \forall i  \\ 
\label{eq:lastRO}
&&u \geq 0;~ v_i \geq 0;~\gamma_i \geq 0;~\mu_i \geq 0;~\sigma_i \geq 0, \forall i. 
\eeqn

This is an MILP problem, which can be efficiently solved using solvers such as Gurobi.
Note that the maximum operator in the average delay constraint (\ref{eq:aveDR}) is removed because at the optimum (i.e., minimum) of the robust problem (\ref{eq:ROobjfinal})-(\ref{eq:lastRO}), the last term in the right hand side of  (\ref{eq:aveDR}) should be maximized so that the  feasibility set is largest.

\subsection{The KKT Conditions }
\label{SPKKTA}

The Lagrangian function of the inner minimization problem 
(\ref{eq:SPIobj})-(\ref{eq:SPce})
in the subproblem \textbf{SP} is:
\beqn
\mathcal{L}(x,\pi_0,\pi_j,\mu,\sigma_i,\xi_{i,0},\xi_{i,j}) = -\sum_{i,j}\xi_{i,j} x_{i,j} - \sum_i x_{i,0} \xi_{i,0}   \nonumber \\ \nonumber
 \beta  \Big( \sum_i d_{i,0} x_{i,0} + \sum_{i,j} d_{i,j} x_{i,j} \Big) + \pi_0 (w \sum_i x_{i,0} - y_0) \\ \nonumber
+ \sum_j \pi_j ( w\sum_i x_{i,j} - y_j) + \sum_i \sigma_i (\lambda_i - x_{i,0} - \sum_j x_{i,j}) \\ \nonumber
+\mu \Big( \sum_i d_{i,0} x_{i,0} + \sum_{i,j} d_{i,j} x_{i,j}  - D^{\sf m} \sum_i \lambda_i \Big) 
\eeqn
The KKT conditions give
\beqn
\label{eq:kktx0}
\frac{\partial L}{\partial x_{i,0}} = \beta d_{i,0} + w \pi_0  + \mu d_{i,0} - \sigma_i - \xi_{i,0} = 0,~\forall i \\
\label{eq:kktxi}
\frac{\partial L}{\partial x_{i,j}} = \beta d_{i,j} + w \pi_j  + \mu d_{i,j} - \sigma_i - \xi_{i,j} = 0,~\forall i,j \\
\label{eq:kktps}
w \sum_i x_{i,0} \leq y_0;~~
  w\sum_i x_{i,j} \leq  y_j,~~ \forall j \\ 
\sum_i d_{i,0} x_{i,0} + \sum_{i,j} d_{i,j} x_{i,j}  \leq  D^{\sf m} \sum_i \lambda_i, 
\eeqn
\beqn
\label{eq:kktpe}
 x_{i,0}  + \sum_j x_{i,j} =  \lambda_i,~ \forall i;
 x_{i,0} \geq 0,~\forall i;~
 x_{i,j} \geq 0,~ \forall i,j \\
 \label{eq:kktfea1}
\pi_0 \geq 0;~~\pi_j \geq 0,~\forall j;~~ \mu \geq 0;~~\sigma_i \geq  0,~\forall i \\
\label{eq:kktfea2}
\xi_{i,0} \geq 0,~\forall i;~~\xi_{i,j} \geq 0,~\forall i,j\\
 \label{eq:kktcs}
  \big(y_0 - w \sum_i x_{i,0} \big)  \pi_0 = 0; 
 \big(y_j - w \sum_i x_{i,j}\big) \pi_j = 0,\forall j  \\ 
 \big(  D^{\sf m} \sum_i \lambda_i - \sum_i d_{i,0} x_{i,0} - \sum_{i,j} d_{i,j} x_{i,j} \big)  \mu = 0 \\
\label{eq:kktce}
x_{i,0} \xi_{i,0} = 0,~~\forall i;~~ 
x_{i,j} \xi_{i,j} = 0,~~\forall i,j,
\eeqn
where (\ref{eq:kktx0})-(\ref{eq:kktxi}) are the stationary conditions, (\ref{eq:kktps})-(\ref{eq:kktpe}) are
the primal feasibility conditions, (\ref{eq:kktfea1})-(\ref{eq:kktfea2}) are the dual feasibility conditions,
 and (\ref{eq:kktcs})-(\ref{eq:kktce}) are complementary slackness conditions.
From (\ref{eq:kktx0})-(\ref{eq:kktxi}), we have:
\beqn
 \beta d_{i,0} + w \pi_0  + \mu d_{i,0} - \sigma_i = \xi_{i,0} \geq 0,~~\forall i \\
 \beta d_{i,j} + w \pi_j  + \mu d_{i,j} - \sigma_i = \xi_{i,j} \geq 0,~~\forall i,j 
\eeqn
From (\ref{eq:kktce}), if $x_{i,0} \geq 0$, then $\xi_{i,0} = 0$ and $\beta d_{i,0} + w \pi_0  + \mu d_{i,0} - \sigma_i = 0$. Thus, $\Big( \beta d_{i,0} + w \pi_0  + \mu d_{i,0} - \sigma_i \Big) x_{i,0} = 0,~ \forall i.$
Similarly, we have $\Big( \beta d_{i,j} + w \pi_j  + \mu d_{i,j} - \sigma_i \Big) x_{i,j} = 0,~ \forall i,j.$
On the other hand, assume $\Big( \beta d_{i,0} + w \pi_0  + \mu d_{i,0} - \sigma_i \Big) x_{i,0} = 0,~ \forall i$. If $x_{i,0} > 0$, then $\beta d_{i,0} + w \pi_0  + \mu d_{i,0} - \sigma_i = 0$, which implies $\xi_{i,0} = 0$. Hence, $x_{i,0} \xi_{i,0} = 0,~\forall i.$ 
Similarly, $\Big( \beta d_{i,j} + w \pi_j  + \mu d_{i,j} - \sigma_i \Big) x_{i,j} = 0,~ \forall i,j$ implies $x_{i,j} \xi_{i,j} = 0,~\forall i,j.$  

Therefore, it is easy to see that the KKT conditions  (\ref{eq:kktx0})-(\ref{eq:kktce}) and 
the set of constraints (\ref{eqe:comps})-(\ref{eqe:compe})
 are equivalent. It is worth noting that we can directly use the  set of KKT conditions  (\ref{eq:kktx0})-(\ref{eq:kktce})
 to solve the subproblem, but it will involve more variables (i.e., $\xi_0$ and $\xi$) compared to solving the subproblem with (\ref{eqe:comps})-(\ref{eqe:compe}).
Furthermore, the set of constraints (\ref{eqe:comps})-(\ref{eqe:compe}) can
be informally and quickly obtained by using duality in linear programming, without explicitly writing down the Lagrangian and KKT conditions.

\subsection{The Complete MILP Formulation for the Subproblem}
\label{MILPSP}
By applying  the Fortuny-Amat  transformation \cite{bigM} to every complimentary constraint in the problem 
(\ref{eq:SPobjf})-(\ref{eq:SPlasteq}), we obtain the equivalent MILP to the  subproblem (\ref{SP}) as follows: 
\beqn
\label{milpSPobj}
\underset{x,\lambda,\pi_0,\pi_j,\mu,\sigma_i,u}{\text{max}}
~\beta  \Bigg\{ \sum_i d_{i,0} x_{i,0} + \sum_{i,j} d_{i,j} x_{i,j} \Bigg\}
\eeqn
subject to
\beqn
0 \leq \beta d_{i,0} + w \pi_0 + \mu d_{i,0} - \sigma_i \leq u_i^0 M_i^0,~~\forall i\\
0 \leq x_{i,0} \leq (1-u_i^0) M_i^0,~~\forall i\\
 0 \leq  \beta d_{i,j} + w\pi_j + \mu d_{i,j} - \sigma_i \leq u_{i,j}^1 M_{i,j}^1,~~\forall i,j\\
 0 \leq x_{i,j} \leq (1-u_{i,j}^1) M_{i,j}^1,~~\forall i,j\\
0 \leq y_0 - w \sum_i x_{i,0} \leq u^2 M^2\\
0 \leq  \pi_0 \leq (1-u^2) M^2\\ 
 0 \leq y_j - w \sum_i x_{i,j} \leq u_j^3 M_j^3,~~\forall j  \\ 
 0 \leq \pi_j \leq (1-u_j^3) M_j^3,~~\forall j  \\ 
 \label{MILPSPeq1}
 0 \leq    D^{\sf m} \sum_i \lambda_i - \sum_i d_{i,0} x_{i,0} - \sum_{i,j} d_{i,j} x_{i,j} \leq u^4 M^4\\
0 \leq  \mu \leq (1-u^4) M^4\\
u_i^0 \in \{0,1\},~\forall i;~ u_{i,j}^1 \in \{0,1\},~\forall i,j \\
u^2,~u^4 \in \{0,1\};~u_j^3 \in \{0,1\}\\
\label{MILPSPeq2}
x_{i,0}  + \sum_j x_{i,j}  = \lambda_i,~~\forall i\\
\label{MILPSPeq3}
\lambda_i = \lambda_i^{\sf f} + g_i \hat{\lambda}_i,~\forall i ;~~
\sum_i t_i \leq \Gamma;~~  t_i \leq 1,~~ \forall i \\
\label{MILPSPeq4}
-t_i + g_i \leq 0,~~ \forall i;~~t_i + g_i \geq 0,~~ \forall i,  
\eeqn
where $u$ represents the set of binary variables $u^0, u^1, u^2$, $u^3, u^4$. Also,  $M_i^0, M_{i,j}^1, M^2, M_j^3, M^4, M_i^5$ are sufficiently large numbers. The value of each $M$ should be large enough to ensure feasibility of the associated constraint. On the other hand, the value of each $M$ should not be too large to enhance the computational speed of the solver. Indeed, the value of each $M$ should be tighten to the limits of parameters and variables in the corresponding constraint. For instance, $M_i^0$ needs to be larger or equal to the maximum value of $x_{i,0}$, which is $\lambda_i$. Thus,  $M_i^0$ can be set to be a small number greater than $\lambda_i$.

Additionally, it is easy to see that, for the maximization problem (\ref{milpSPobj})--(\ref{MILPSPeq4}), $g_i$ should be nonnegative for every $i$  to maximize the objective function (see constraints  (\ref{MILPSPeq1}),   (\ref{MILPSPeq2}),  (\ref{MILPSPeq3}), (\ref{MILPSPeq4})).
Hence, we can simplify (\ref{MILPSPeq3})-(\ref{MILPSPeq4}) as
\beqn
\lambda_i = \lambda_i^{\sf f} + t_i \hat{\lambda}_i,~\forall i;~ \sum_i t_i \leq \Gamma;~ 0 \leq  t_i \leq 1,~ \forall i.
\eeqn

\subsection{Convergence of Algorithm 1}
\label{proofcon}

\textbf{Algorithm}
\ref{AROalg} converges to the optimal value of the original two-stage robust problem (\ref{AROori}) in $O(K)$ iterations. Indeed, this can be shown by contradiction that any repeated $\lambda^*$ implies $LB = UB$.
Specifically, assume  $(z^*, y^*, \eta^* )$ is the optimal solution to
MP (\ref{algobjaro}), $(\lambda^*, x^*)$ is the optimal solution to the SP in iteration $k$, and $\lambda^*$ appears in a previous iteration. From step 4 of \textbf{Algorithm}
\ref{AROalg}, we have:
$UB \leq  \sum_j h_j z_j^* + \sum_j p_j y_j^* + p_0  y_0^* + \beta  \Big( \sum_i d_{i,0} x_{i,0}^* 
+ \sum_{i,j} d_{i,j} x_{i,j}^* \Big) = RHS.$ Now, since $\lambda^*$ appears in a previous iteration, the MP in iteration $k+1$ is identical to  the MP in iteration $k$. Hence,  $(z^*, y^*, \eta^* )$ is also the optimal solution to the MP in iteration $k+1$. 
We have: $LB \geq  \sum_j h_j z_j^* + \sum_j p_j y_j^* + p_0  y_0^* + \eta^*.$
Since the extreme scenario $\lambda^*$ has already been identified and related constraints are added to the MP before iteration $k$, we have: $\eta^* \geq  \beta  \Big( \sum_i d_{i,0} x_{i,0}^* 
+ \sum_{i,j} d_{i,j} x_{i,j}^* \Big)$. Thus, $LB \geq RHS \geq UB$, which implies $LB = UB$. 
As a result, \textbf{Algorithm}
\ref{AROalg} converges in a finite number of iterations. Typically, the algorithm converges within a few iterations as shown in the numerical results.

\subsection{Stochastic Formulation}
\label{stoc}
The stochastic formulation of the service placement and sizing problem is the following MILP problem:
\beqn
\label{eq:stocObj}
 \underset{x,y,z}{\text{min}}~  \sum_j f_j (1 - z_j^0) z_j + \sum_j s_j z_j +  p_0  y_0  +  \sum_j p_j y_j    \nonumber \\ 
 + \sum_{\xi} \eta^{\xi}  \Bigg\{ \beta \big( \sum_i d_{i,0} x_{i,0}^\xi + \sum_{i,j} d_{i,j} x_{i,j}^\xi \big) \Bigg\}
\eeqn
subject to 
\beqn
\sum_j z_j \geq r^{\sf min} \\ 
y_j \leq z_j C_j, ~~ \forall j \\
\sum_j f_j (1 - z_j^0) z_j + \sum_j s_j z_j + p_0  y_0  + \sum_j p_j y_j  \leq B \\
w \sum_i x_{i,0}^\xi \leq y_0;~~
w \sum_i x_{i,j}^\xi \leq y_j, ~~ \forall j, \xi \\
x_{i,0}^\xi  +  \sum_j x_{i,j}^\xi 
 = \lambda_i^\xi, ~~ \forall i, \xi \\
\sum_i d_{i,0} x_{i,0}^\xi + \sum_{i,j} d_{i,j} x_{i,j}^\xi  \leq \sum_i \lambda_i^w D^{\sf m}, ~\forall \xi  \\  
z_j \in \{0,~1\};~~y_0 \geq 0;~~y_j \geq 0,~\forall j \\ 
\label{eq:stocLast} 
x_{i,0}^\xi \geq 0,~\forall i;~~x_{i,j}^\xi \geq 0,~ \forall i, j, \xi.
\eeqn
Note that $\xi$ is the scenario index and $\eta^\xi$ is the probability of scenario $\xi$, where $\lambda_i^\xi$ is the demand at AP $i$ in scenario $\xi$. Also, $x_{i,j}^\xi$ is the workload allocated from AP $i$ to EN $j$   and $x_{i,0}^\xi$ is the workload allocated from AP $i$ to the cloud in scenario $\xi$.
The meaning of the variables and constraints in the stochastic model is similar to those in the deterministic model in  \textit{Appendix} \ref{DetermiA}. The main difference is that we consider a set of scenarios in the stochastic model to express the uncertainty.

The stochastic model assumes that $\eta^{\xi}$ and the demand at every AP in each scenario $\xi$ are known and  given as input to the  problem (\ref{eq:stocObj})-(\ref{eq:stocLast}). Thus, the stochastic approach needs to know the exact probability distribution of the uncertainties to have a good performance. This information is hard to obtain in practice. Furthermore, the uncertainty realization may not follow the historical pattern (i.e., future can be different from the past). The proposed robust methods do not need the probability distribution information. Furthermore, the  objective of a stochastic model is to optimize the ``average'' or ``expected" system performance
over all the scenarios, while the goal of a robust model is to optimize the ``worst-case'' performance. Thus, the design objectives of the stochastic and robust approaches are different. The optimal solution obtained from a stochastic model can perform badly in worse-case scenarios, especially for long-tail distributions.


In the stochastic model (\ref{eq:stocObj})-(\ref{eq:stocLast}), similar to the ARO model, the service placement and sizing decisions are made in the first stage before the uncertainty is revealed. Given the first-stage decisions, the workload allocation decision is determined in the second stage after knowing the actual demand realization. 
Since the actual demand can be different from the generated scenarios (which can be understood as training scenarios), we need to evaluate the performance of the stochastic model in the actual operation stage over a set of testing scenarios. As in the deterministic model, the procured resources in the first-stage of the stochastic model may not be sufficient to serve the actual demand and the SP may need to drop some requests.

To compare the performance of the stochastic approach with the deterministic and robust approaches, we generate 1000 training demand scenarios to represent the demand uncertainty. The  scenario set is then used as the input to the stochastic problem (\ref{eq:stocObj})--(\ref{eq:stocLast}) to determine the optimal placement and sizing solution in the first stage. Then, similar to Section \ref{operationCompare} in the main manuscript, we generate 100 testing demand scenarios to compare the performance of different schemes.

\begin{figure}[ht!]
	\centering
		\includegraphics[width=0.40\textwidth,height=0.18\textheight]{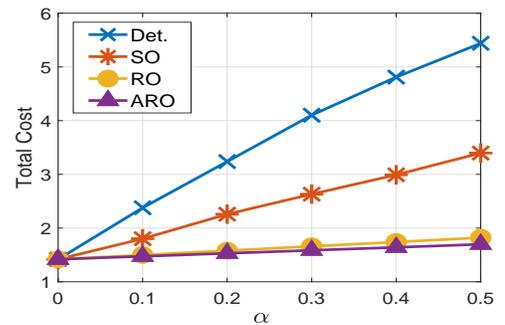}
			\caption{Worst-case cost comparison in the  operation stage}
	\label{DSRAR}
\end{figure}

Fig.~\ref{DSRAR} illustrate the performance of the deterministic, stochastic (SO), RO, ARO approaches.
To evaluate the performance of the stochastic approach, we need to assume that we know the exact distribution of the demand across different areas to generate a set of training demand scenarios. Since the exact distribution is hard to obtain and also future data can be different from the historical pattern, the actual performance of the stochastic model can be worse than the reported performance in Fig. \ref{DSRAR} where we assume to  know the exact probability distribution of the demand.
For the results in Fig. \ref{DSRAR}, the demand distribution at the APs is assumed to follow a multivariate normal distribution. 
We have also run simulations with other distributions and obtained similar trends and observations on the performance comparison of the four approaches. The SO solution outperforms the deterministic one since it captures the uncertainty, but it is still much worse than the proposed robust models.

\end{document}